\documentclass[journal]{IEEEtran}
\usepackage{cite}
\usepackage{amsmath,amssymb,amsfonts}
\usepackage{graphicx}
\usepackage{textcomp}
\usepackage{booktabs}
\usepackage{array}
\usepackage{url}
\usepackage{xurl}        
\usepackage{bm}

\newenvironment{termdesc}
  {\begin{list}{}%
    {\setlength{\leftmargin}{1em}\setlength{\itemindent}{-1em}%
     \setlength{\labelsep}{0pt}\setlength{\labelwidth}{0pt}%
     \setlength{\itemsep}{0.4ex}\setlength{\parsep}{0pt}\setlength{\topsep}{0.5ex}%
     }}
  {\end{list}}

\begin{document}
\title{A Verifier-Centric Conceptual Model\\ for Digital Credential Ecosystems\\[2mm]{\large\normalfont Decomposing Verification into Establishment, Acceptance, and Materials Acquisition}}

\author{Shigeya~Suzuki,~\IEEEmembership{Member,~IEEE,}
and~Ryosuke~Abe%
\thanks{S. Suzuki is with the Global Research Institute, Keio University, Tokyo, Japan (e-mail: shigeya@wide.ad.jp).}%
\thanks{R. Abe is with the Japan Advanced Institute of Science and Technology, Ishikawa, Japan (e-mail: ryosuke@jaist.ac.jp).}%
\thanks{Preprint. This work has been published in IEEE Access, vol.~14, pp.~146300--146323, 2026, doi: 10.1109/ACCESS.2026.3735288. This preprint is typeset with IEEEtran and does not reflect the copy-editing and typesetting of the published version; wording, house style, and the formatting of references differ in places. Please cite the published version.}}

\markboth{Suzuki \textit{et al.}: A Verifier-Centric Conceptual Model for Digital Credential Ecosystems}{Suzuki \textit{et al.}: A Verifier-Centric Conceptual Model for Digital Credential Ecosystems}

\maketitle

\begin{abstract}
Digital credential ecosystems increasingly combine multiple standards. Because implementations have evolved independently across jurisdictions and application domains, systems described under the common label ``digital credential'' often remain mutually non-interoperable. Conventional element-by-element comparisons of identifiers, data models, credential formats, protocols, and signature algorithms do not explain why interoperability fails even when stacks share a data model, nor do they identify what a verifier must obtain, and what it must trust, before accepting a credential. We present a verifier-centric conceptual model built on two decompositions. The first separates credential processing into signature verification (L1), semantic interpretation (L2), and validation (L3), and models the supporting materials through two orthogonal planes: Constitution, which captures ecosystem-level arrangements and trust declarations, and Logistics, which captures how verification materials are stored and delivered; the Shinken framework makes trust assumptions explicit across all five functions. The second characterizes where each function may be placed along three dimensions (placement, timing, and disclosure). From the condition of being verifiable, the model derives seven consequences, distinguished as definitional corollaries, operational implications, and design trade-offs. Applying the model to four learner-credential stacks and to existing ecosystems including authentication federations, we show that it explains interoperability failures, verifier-side burden, offline verifiability, privacy implications, and terminological ambiguities that element-wise comparison leaves unresolved.
\end{abstract}


\section{Introduction}
\label{sec:intro}
\IEEEPARstart{T}{his} paper treats the activity spanning the issuance and verification of digital credentials as a single ecosystem and presents a conceptual model of it. The discussion assumes the Issuer--Holder--Verifier (IHV) model; the activities of these three parties, together with the shared arrangements that support them, combine to form one ecosystem.

This ecosystem is not closed under a single standard or a single operating authority. In practice, multiple technical stacks that do not interoperate coexist under the common label ``digital credential.'' Juxtaposing technical elements such as identifier schemes, data models, and signature algorithms offers no handle for answering why interoperability does not hold, or what must be present for a verifier to verify. The position of this paper is that interoperability is not a matter of format compatibility: it is the condition that a verifier can obtain the necessary verification materials and reach an acceptance decision under explicitly stated assumptions. Such questions can be discussed only after introducing a model of the ecosystem's structure: what is shared where, how it reaches the verifier, and how the verifier's role is divided.

We therefore replace the juxtaposition of technical elements with a conceptual model that decomposes the activities constituting the ecosystem into functional layers and planes and captures how their combination enables verification. The model prescribes no particular standard; its purpose is to provide a common frame for interpreting and comparing existing standards and systems across stacks, and for evaluating designs.

The paper is organized as follows. We first present, as observations, phenomena that the existing comparison axes fail to capture, and distill them into a problem statement. After reviewing related work, the conceptual-model section presents a model responding to those points: three layers and two planes, the Shinken framework that makes the introduction of trust explicit, and three dimensions of role decomposition. The application-and-evaluation section applies it to existing stacks in the learner-credential domain and to related ecosystems, confirming its interpretive power. We conclude with a summary. The appendix gives an overview of Trusted Web and the Shinken model, sources of this model that are available only in Japanese.

\subsection{Observations: Four Viewpoints}
The use of digital credentials advances through combinations of multiple standards and specifications, including W3C VCDM~\cite{vcdm2}, ISO/IEC 18013-5 (mdoc)~\cite{iso18013}, SD-JWT VC~\cite{sdjwtvc}, OpenID for Verifiable Credentials~\cite{oid4vci,oid4vp}, eIDAS~2.0~\cite{eidas2}, and Open Badges 3.0~\cite{ob3}. Standards grounded in the issue--hold--verify IHV model coexist with those built on the pre-existing X.509 PKI and on federated identity infrastructures, and with platform-proprietary schemes.

Because design and implementation have proceeded independently in each domain, systems described under the common label ``digital credential'' are implemented as multiple stacks that do not interoperate. The phenomenon is especially pronounced in the learner-credential domain, where at least four kinds of stacks (the 1EdTech family, the unprofiled use of W3C DID/VC~\cite{did1,vcdm2}, the European EDC/ELM/EUDI-Wallet, and platform-proprietary ones) and more than six silos formed by their combinations can be observed~\cite{edcredi}.

Examining these silos, we identify four viewpoints.

\subsubsection{Viewpoint 1: The existing comparison axes and their limits}
In discussing differences among standards, it is conventional to compare technical elements such as identifier schemes, transport protocols, data models, credential formats, and signature algorithms. Such a comparison table is useful as an overview of standards and implementations.

This comparison axis, however, has the following limitations:

\begin{itemize}
\item It does not explain why stacks fail to interoperate even when they adopt the same data model.
\item It offers no comparable terms in which to organize the different mechanisms by which stacks confirm issuer legitimacy.
\item It does not answer, across stacks, what must be present for a verifier to receive and verify a credential, or by what mechanism it comes to be present.
\end{itemize}
\subsubsection{Viewpoint 2: A tendency for the discussion to center on the issuer}
Existing discussion tends to concentrate on issuer-side design (which standard to issue with, which data model to choose, under which governance to operate), leaving verifier-side design (on what premises the verifying system runs, what it trusts, what must be present for verification to hold) comparatively underexamined.

This bias is coupled with a terminological confusion in which terms such as registry, trust framework, trust list, and issuer metadata are used across stacks while pointing at different objects. Because discussion proceeds without agreement on what the terms denote, consensus is hard to accumulate.

\subsubsection{Viewpoint 3: The missing discussion of the storage and supply of issuer information}
As a third viewpoint, stacks differ greatly in how far they specify the mechanism by which the information needed for credential verification is actually stored and reaches the verifying system.

In the European EDC/ELM~\cite{elm}/EUDI-Wallet~\cite{eudiarfv2} stack, the storage and supply paths of issuer information are specified at the protocol level, so that the information needed at verification time is systematically assembled. Other stacks lack specification from this viewpoint and leave the matter to implementation or to implicit premises. A frame that treats this viewpoint as an object of discussion has been lacking.

\subsubsection{Viewpoint 4: The missing discussion of the division of the verifier's role}
As a fourth viewpoint, who divides---and when and where---the work the verifier must perform to reach trust establishment is not discussed explicitly. Even for the same work of tracing a trust chain, the verifier-side burden, privacy, and network cost differ greatly depending on whether the verifier performs it itself, delegates it to an external resolver, has the issuer bundle the trust chain into the credential in advance, or has a pre-aggregator aggregate and distribute the whole.

Judgments such as ``Federation is too heavy'' or ``aggregated distribution is centralized'' rest on this division of roles; yet because no frame treats the division of roles as an independent object of discussion, the grounds for such judgments remain unorganized.

\subsection{The Problem This Paper Addresses}
\label{sec:problem}
The phenomena captured by the four viewpoints above are what juxtaposed comparison of technical elements has overlooked. This paper discusses these points along the following four axes, which become the pillars of the analysis the model performs from the conceptual-model section onward:

\begin{itemize}
\item how the set of issuers is structured and what is shared as a community;
\item how the content shared in the community, and the credential itself, reach the verifying system;
\item from the verifier's viewpoint, what is built into the verification logic and what is fetched at runtime; and
\item who divides the verifier's role, and when and where.
\end{itemize}
In what follows, this paper presents a conceptual model for organizing these points. As a starting point, we rely on the abstract model of a community that uses digital signatures, presented in the Trusted Web White Paper ver.3.0~\cite{tw3}, and on our own research results that develop this notion~\cite{claimval,shinken}.

\subsection{Contributions}
The contributions of this paper are as follows:

\begin{itemize}
\item A functional decomposition of credential verification into three layers that carry the establishment and acceptance of verification---Signature Verification (L1), Semantic Interpretation (L2), and Validation (L3)---and two orthogonal planes that carry the acquisition of verification materials---Constitution and Logistics. The Shinken framework, a formalization of trust, is then applied across these five functions, making the introduction of trust explicit (the conceptual-model section).
\item A decomposition of roles along three dimensions (placement, timing, disclosure). It makes the verifier's role an explicit object of analysis and yields seven consequences drawn from the condition of being verifiable (Sections~\ref{sec:II_5},~\ref{sec:II_6_conseq}).
\item A cross-sectional application to four learner-credential stacks and to existing ecosystems including authentication systems. We show that the model resolves well-known terminological confusions (the application-and-evaluation section).
\end{itemize}
\section{Related Work}
\label{sec:related}
The IHV model mentioned at the outset specifies credential formats, identifier abstractions, and issuance/presentation protocols, but largely leaves both the constitution of the community and the supply of verification materials to implementations. The European EDC/ELM~\cite{elm}/EUDI-Wallet~\cite{eudiarfv2} stack is a rare example in which the storage and supply paths of issuer information are specified at the protocol level. The governance aspect is addressed by the Trust over IP dual-stack model~\cite{toip,toipstack} and the Open Identity Exchange (OIX) trust-framework methodology~\cite{oix}; both are normative, methodological frameworks prescribing how an ecosystem should be built and governed. SIDI Hub (Sustainable and Interoperable Digital Identity Hub)~\cite{sidihub}, a community activity involving standards bodies and governments, addresses cross-border interoperability of digital identity through use-case identification, minimum interoperability requirements, and the mapping of trust frameworks across jurisdictions; it frames the problem as connecting arrangements among communities, not as the decomposition of verification or the supply structure of verification materials. To our knowledge, no existing framework treats either the storage-and-supply plane or the division of the verifier's role as an independent object of analysis.

The OpenID4VC High Assurance Interoperability Profile (HAIP)~\cite{haip} secures interoperability at the level of a profile: combining OID4VCI, OID4VP, SD-JWT VC, and mdoc, it fixes each specification's optionalities to define interoperability requirements among issuers, wallets, and verifiers. The standardization practice that individual specifications alone do not suffice for cross-stack interoperability, and that a profile fixing the choices is required, is consistent with the background of Viewpoint 1. HAIP itself, however, explicitly places trust management, such as the authorization of issuers, outside its scope---including the methods for establishing trust and obtaining root certificates; that is, it does not define what this paper calls the Constitution as a whole---and provides no analytical frame for the acquisition structure of verification materials or the division of the verifier's role.

For the publication and discovery of trust schemes, the lineage of LIGHTest~\cite{lightest} and its successor TRAIN~\cite{train} uses DNS as the delivery path, and its application to cross-domain decentralized-identity infrastructures continues in recent work~\cite{traineudi}. Krul et al.'s SoK~\cite{ssisok} systematizes the requirements and assumptions of trust in SSI. As specifications that prescribe trust declarations themselves as a data model, DIF's Trust Establishment~\cite{difte} and, building on it, Credential Trust Establishment (CTE)~\cite{difcte} address the authorization of issuers: CTE describes participants' roles and authority as a governance document that an authority signs and publishes, supplying materials for judging whether an issuer may be trusted. Trust Establishment defines only the data model of trust documents, leaving their integrity protection, format, publication, and discovery to profiles; CTE likewise leaves the specifics of signing and publishing the declarations to separately defined profiles. These efforts share subject matter with the Constitution and Logistics planes in that they treat trust schemes, trust declarations, and their delivery paths; their concern, however, lies in describing, publishing, discovering, and systematizing them, not in connecting them to verifier-side processing.

The model presented here builds on our own results. The Trusted Web White Paper ver.3.0~\cite{tw3} presents a model of an ecosystem that uses digital-signature technology---how signed data can be handled within a community. The Claim Validation model~\cite{claimval} separates verification (confirmation of integrity and data origin, independent of subject and context) from validation (acceptance against the verifier's criteria, dependent on the verifier's intent). The Shinken model~\cite{shinken} is an abstract model for implementing validation as an information system; it formalizes trust as assuming some propositions true within a logic of verification that can contain propositions not decidable from computational evidence alone. Making this introduction of trust explicit yields determinism and transparency, and localizes which assumption to revise when one breaks. This paper adopts these as the core of the model's design, refining the Claim Validation model's verification into L1 and L2 and extending the Shinken model's formalization of trust into the Shinken framework that runs across all three layers and two planes. The Shinken framework differs in scope from formal trust models in the security literature, such as models of public-key infrastructures~\cite{maurerpki} and calculi of trust and reputation~\cite{trustsurvey}: those lines of work quantify trust or derive new trust relations from existing ones, whereas the Shinken framework treats trust as a non-derivable act of acceptance, and confines itself to making that acceptance explicit and attributable. An application of this line of work to agent-oriented data spaces, combining decentralized discovery mechanisms with machine-executable validation policies, is given in~\cite{vaia}. An overview of Trusted Web and the Shinken model, which are available only in Japanese, is given in Appendix~\ref{app:C}.

\subsection{Difference from Existing Work}
The novelty of this paper lies not in proposing a new governance framework but in an analytical model that connects verifier-side reasoning (L1--L3) with the acquisition of verification materials (Constitution and Logistics), and further treats the freedom of placement of roles (the three dimensions) and the introduction of trust (the Shinken framework) in a single vocabulary. The model does not replace these existing efforts; it connects the scopes they address to verifier-side reasoning and to the structure of acquisition. Table~\ref{tab:related} maps the scope of the existing efforts discussed in this section onto the components of this model. Each effort addresses a subset of these; to our knowledge, no existing framework treats them jointly as a single object of analysis. Here, ``No'' does not mean that the work is silent on governance or trust frameworks, but that it does not treat them as an object of analysis connected to verifier-side reasoning (L1--L3) and role placement in the sense of this paper.

\begin{table*}[htbp]
\caption{Scope of related work mapped onto the components of this model (Yes: treated as an object of analysis; Partial: partially; No: not, in the sense of this paper)}
\label{tab:related}
\centering
\footnotesize
\begin{tabular}{@{}p{3.0cm}p{2.2cm}p{2.3cm}p{2.2cm}p{2.0cm}p{2.6cm}@{}}
\hline\hline
\textbf{Work} & \textbf{Constitution plane} & \textbf{Logistics plane} & \textbf{Connection to L3 Validation} & \textbf{Role placement} & \textbf{Trust formalization} \\ \hline
ToIP dual-stack & Yes (normative) & No & No & No & No \\
OIX trust-framework methodology & Yes (normative) & No & No & No & No \\
SIDI Hub & Partial (inter-community connection) & No & No & No & No \\
LIGHTest/TRAIN & Partial (content of trust schemes) & Yes (publication, discovery, delivery) & No & No & No \\
DIF TE/CTE & Partial (trust-declaration data model) & No & No & No & No \\
Krul et al.'s SoK & Partial & No & No & No & Partial (requirements and assumptions) \\
This paper & Yes & Yes & Yes & Yes & Yes \\
\hline
\end{tabular}
\end{table*}
\section{The Conceptual Model}
\label{sec:model}
\subsection{Overview of the Model}
\label{sec:II_1}
This section develops a conceptual model of the digital credential ecosystem, starting from existing discussions of digital credential models. The distinctions introduced here are not meant to replace the terminology of individual specifications; they are analytical distinctions for comparing different stacks from the same standpoint.

\subsubsection{Base concept: credentials and signed data}
A credential in this model is data in which an issuer states, under its signature, a claim about some subject, represented as cryptographically signed data. We call this signed data and describe it with the following vocabulary.

\emph{Payload}: the body of data over which the signature is computed. It holds the content (the claim) that the credential asserts.

\emph{Key}: the key used to produce and verify a signature. In public-key cryptography, a signing key (the private key) and a verification key (the public key) form a pair. Hereafter ``key'' denotes this pair; this paper deals chiefly with the verification key. Signed data usually carries either the verification key itself or a clue for discovering it.

\emph{Signer}: the party that controls the signing key and applies a signature to the payload.

\emph{Controller of the signing key}: the party that keeps the signing key---the secret part of the pair---under its control. Usually identical to the signer.

\emph{Signature}: the cryptographic evidence the signer applies to the payload with the signing key. With the verification key one can confirm that the payload has not been tampered with and that the controller of the signing key signed it.

A signature establishes only two facts: that the controller of the signing key signed the payload, and that the payload is identical to what was signed. The intent behind the signature, and whether the claim may be accepted, are not settled by the signature itself; the model assigns these to two further layers, semantic interpretation and validation.

\subsubsection{Base concept: community}
At the base of the model lies the notion of a community: a group that shares common arrangements concerning digital credentials, with various parties involved as stakeholders. The arrangements cover which data models and protocols to use, whom to recognize as issuers, and what to take as the origin of trust; sharing the technical stack is a particularly important part. The ``coexistence of multiple non-interoperable technical stacks'' observed in the introduction is, seen this way, precisely the coexistence of multiple communities, each sharing different arrangements.

\begin{figure}[tb]
\centering
\includegraphics[width=0.95\columnwidth]{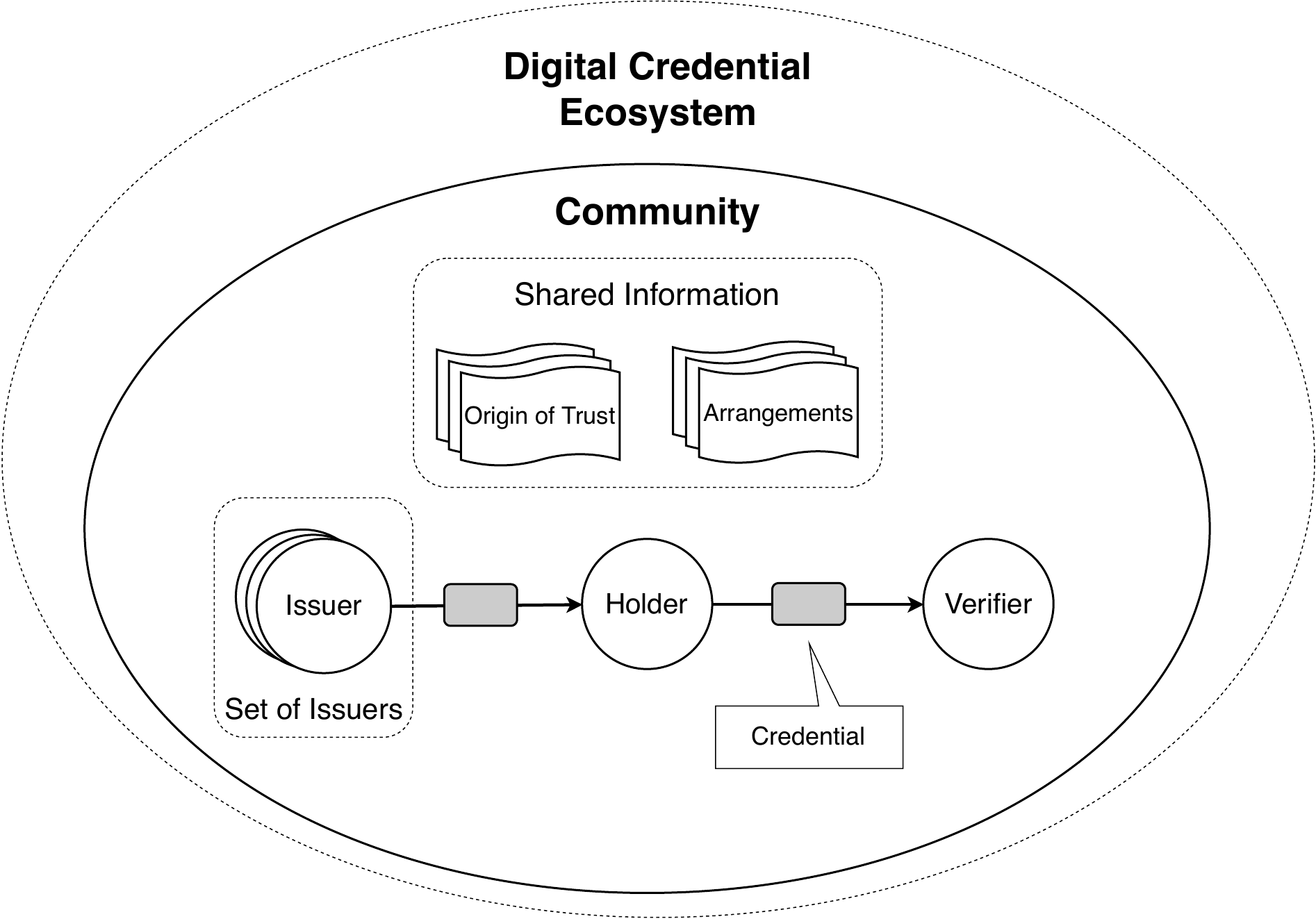}
\caption{Overview of the digital credential ecosystem}
\label{fig:ecosystem}
\end{figure}

A community holds together because issuer, holder, and verifier share its arrangements: the issuer issues credentials, the holder carries them, and the verifier verifies by relying on those arrangements, in particular the origin of trust. All three are members, with differing modes of involvement. Membership here requires neither legal personality nor a formal membership system. Whom to recognize as an issuer is prescribed by the arrangements, and that collection forms the issuer set; although community and issuer set may look identical from some viewpoints, this model treats the issuer set as one constituent element among the arrangements---the policy, identifiers and namespace, the issuer set, the origin of trust, and other metadata. How these arrangements are constituted, and how a community comes to be established, are discussed in Section~\ref{sec:II_3_1} (the Constitution plane); the overall picture of the ecosystem is shown in Fig.~\ref{fig:ecosystem}.

A community is not a closed group: a verifier need not handle only the credentials of the community to which it belongs. By accepting the arrangements of an external community, including its origin of trust, the verifier can verify and accept a credential issued in that community. The community boundary can thus be crossed depending on which community's arrangements the verifier accepts; such connection among coexisting communities remains a topic for further development.

\subsubsection{Contour of the model: decomposition of function and decomposition of placement}
The model organizes the four axes set out in the problem statement into two groups of functions.

The first group is the establishment and acceptance of verification: by what logic a single credential comes to be verified and accepted. It comprises three layers---from the bottom, signature verification (L1), semantic interpretation (L2), and validation (L3)---where an upper layer holds only once the lower one holds. This dependence is logical: it does not require that the processing be executed in this order at runtime, nor that the layers be placed in the same location. This section discusses a single credential; Section~\ref{sec:II_7} generalizes to combinations of multiple credentials.

\begin{figure}[tb]
\centering
\includegraphics[width=0.95\columnwidth]{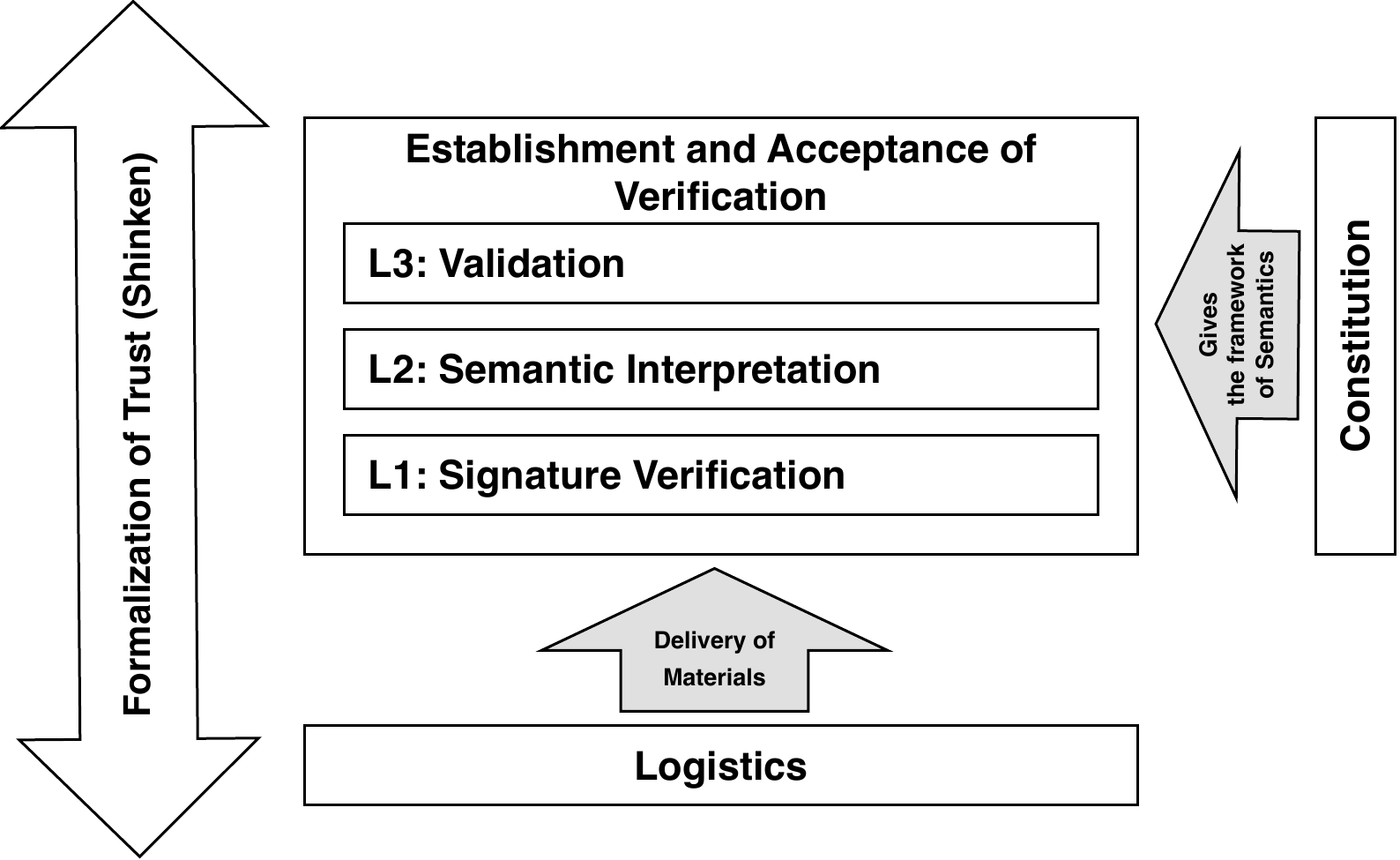}
\caption{The model of three layers and two planes}
\label{fig:layers}
\end{figure}

The second group is the acquisition of the materials that support verification: from where and how the materials needed for the establishment and acceptance of verification are assembled. It comprises two planes: a Logistics plane, handling the delivery path of verification materials, and a Constitution plane, providing, as part of those materials, the arrangements a community shares. The two planes are orthogonal, standing to the three layers in the relation of carrying materials (Logistics) and giving the frame of meaning (Constitution). From the supply side the same flow is the supply of verification materials; we use the verifier-side word \emph{acquisition} to align with origin selection (Section~\ref{sec:II_4}). Fig.~\ref{fig:layers} shows the configuration of the five functions. Table~\ref{tab:functions} summarizes what each function handles and its key property.

\begin{table}[htbp]
\caption{Summary of the five functions. The Shinken framework applies to all five functions.}
\label{tab:functions}
\centering
\footnotesize
\setlength{\tabcolsep}{4pt}
\begin{tabular}{@{}p{0.24\columnwidth}@{\hspace{4pt}}p{0.34\columnwidth}@{\hspace{4pt}}p{0.32\columnwidth}@{}}
\hline\hline
\textbf{Function} & \textbf{What it handles} & \textbf{Key property} \\ \hline
L1 Signature Verification & Confirmation of integrity and data origin & Independent of the verifier and of context \\
L2 Semantic Interpretation & Interpretation of claims per the vocabulary or schema & Depends on the shared vocabulary \\
L3 Validation & Acceptance decision against the validation policy & Depends on the verifier's purpose and context \\
Constitution & The community's arrangements and its composition and establishment & Gives the layers their interpretive frame \\
Logistics & Paths by which verification materials reach the verifier (five elements) & Carries materials to the layers \\
\hline
\end{tabular}
\end{table}
All five functions are interpreted end to end through the formalization of trust, which makes explicit which function, or which verification material, has trust introduced into it. The method is due to our prior work ``Shinken''~\cite{shinken}, which organizes as the introduction of trust the moment of assuming-true a proposition that is not decidable by the verifying system from available computational evidence alone; in this model it belongs to no single function but provides a viewpoint cutting across all of them. Shinken is detailed in Section~\ref{sec:II_4} after the functions are described; Appendix~\ref{app:C} outlines the original argument.

The three model elements that carry the establishment of verification are called ``layers,'' and the two that carry the acquisition of verification materials are called ``planes,'' referred to by individual names (signature verification, semantic interpretation, and so on) as needed. The layers point at phases of the act of verification and stack: an upper layer holds only once the lower one holds. The planes point at the structure of a community's constitution and of information flow; without stacking order, they cut across all three layers, supplying verification materials. Distinguishing the words layer and plane makes this heterogeneity explicit in the terminology itself.

For the plane that expresses a community's constitution, the model uses the word Constitution; the reason for this choice is discussed in Section~\ref{sec:II_3_1}.

The model is thus built on two decompositions. The first, the decomposition of function described above, divides the activity surrounding verification into five functions. The second, the decomposition of placement, treats the work these functions carry---excluding the non-delegable selection of assumptions (\ref{sec:II_4})---as a freedom of placement along three dimensions (where the function is executed, when the inference is raised, and how far it may be disclosed) (\ref{sec:II_5}). In what follows, \ref{sec:II_2} and \ref{sec:II_3} present the five functions, \ref{sec:II_4} the formalization of trust acting on all of them, and \ref{sec:II_5} the decomposition of placement; \ref{sec:II_6} then synthesizes the two decompositions, and \ref{sec:II_6_conseq} draws the consequences of being verifiable.

\subsection{Establishment and Acceptance of Verification (L1--L3)}
\label{sec:II_2}
\subsubsection{L1 Signature Verification}
\label{sec:II_2_1}
Signed data consists of three elements: the verification key (or a reference to it), the payload, and the signature. By verifying the signature with the verification key, the recipient confirms two things: that the payload has not been tampered with after signing (integrity), and that the party controlling the signing key bound to the verification key applied the signature (confirmation of data origin). Strictly, the cryptographic computation itself returns no more than ``this signature is valid for this payload under this verification key''; the latter confirmation---attribution to the controller of the signing key---holds by depending on verification materials showing the binding between the verification key and its controller, and on the assumption taking that binding to be true (Section~\ref{sec:II_4}).

This confirmation is deterministic: given the same signed data and the same verification key, the result is the same whoever verifies, independent of the verifier's intent or context. If the result changes when different verification keys or binding information are obtained, that is not because the cryptographic computation depends on context but because the inputs differ. Schemes using selective disclosure or zero-knowledge proofs are also positioned in the same framework, with L1 returning a verification result according to the proof scheme in question.

Signature verification itself is a cryptographic operation completing in the verifying system's local computation, provided the full set of signed data from the trust anchor to the target credential, with each signature's verification key, is present. From where that set is acquired is a separate matter, generally handled as an external dependency under the acquisition of verification materials (Section~\ref{sec:II_3}); where the computation runs is a matter of placement, discussed under the decomposition of roles (Section~\ref{sec:II_5}). What can be delegated is the materials of verification and their processing; what cannot is the determination of the criteria of what to accept---the validation policy and the selection of assumptions (Section~\ref{sec:II_4}). Once the criteria are fixed, the execution of validation (L3) may itself be entrusted to another; but if another party fixes those criteria in its place, the verifying party in question can no longer be said to accept.

The party in the IHV model is a composite of three roles: signature verifier, interpreter, and validator. In this paper ``verifier'' denotes that composite party; the role that bears L1 alone is named the \emph{signature verifier}, avoiding confusion between the party and the role, and we write \emph{the verifying party} where the contrast with the individual roles needs emphasis. The signature verifier confirms integrity and data origin over signed data and---to repeat---does not enter into what the payload means or whether it may be accepted. Table~\ref{tab:roles} maps the three roles to the terminology of the models this paper builds on.

\begin{table}[htbp]
\caption{Correspondence of the verifier's three roles to the terminology of the referenced models (each referenced model treats the verifier as a single role)}
\label{tab:roles}
\centering
\footnotesize
\setlength{\tabcolsep}{3pt}
\begin{tabular}{@{}l@{\hspace{5pt}}l@{\hspace{5pt}}p{0.26\columnwidth}@{\hspace{4pt}}p{0.16\columnwidth}@{}}
\hline\hline
\textbf{This paper} & \textbf{Layer carried} & \textbf{Claim Validation model} & \textbf{Shinken model} \\ \hline
signature verifier & L1 Signature Verification & Validator & Verifier \\
interpreter & L2 Semantic Interpretation & (same) & (same) \\
validator & L3 Validation & (same) & (same) \\
\hline
\end{tabular}
\end{table}
\subsubsection{L2 Semantic Interpretation}
\label{sec:II_2_2}
Success at signature verification (L1) confirms only that ``the controller of this signing key signed this payload''; what the payload asserts is not yet settled. Signature verification guarantees the origin of a byte sequence but does not give it meaning.

Semantic interpretation is the activity of interpreting what a claim asserts, according to the vocabulary and schema on which the payload relies. For a learner credential, for instance, one reads off, against the payload's vocabulary definitions, which institution awarded which degree or title to which learner, and by which criteria.

The flexibility of semantic expression depends on the descriptive power of the vocabulary used. An extensible framework such as JSON-LD lets a claim's meaning be described flexibly, referring to external vocabularies and fitting existing credentials; the learner domain has vocabularies that describe credential meaning in detail, such as CTDL (Credential Transparency Description Language)~\cite{ctdl}, built on JSON-LD. Used appropriately, such vocabularies can improve interoperability.

The higher the descriptive power, however, the heavier the burden of interpretation: the greater the cost of discovering and fetching the external vocabularies interpretation requires, expanding schemas, and confirming semantic consistency. What tends to be overlooked is the verifier-side burden: having the verifier perform full semantic interpretation of a domain's credentials entirely on its own would be heavy; keeping a vocabulary referable to secure interoperability and having every verifier resolve the whole of it by itself are different matters.

From these viewpoints, semantic interpretation can be seen as a layer with freedom of placement; it need not be completed by the verifier alone. Three divisions of roles, for instance, are conceivable:

\emph{The verifier resolves everything itself}: it obtains all needed information, fully resolves the vocabulary, and interprets the payload's meaning entirely by itself.

\emph{Delegating interpretation}: it entrusts part (or all) of vocabulary resolution and semantic interpretation to an external party and receives the result.

\emph{Limiting the scope of interpretation and resolving details when needed}: it refers only to the parts needed, handling the rest uninterpreted, and refers to details later as needed.

These divisions of roles differ greatly in the verifier's burden, the cost of network dependence, and the dependency relations. Where to place semantic interpretation is an instance of the placement problem in the decomposition of roles (Section~\ref{sec:II_5}); among the layers of the establishment of verification, its freedom of placement is the most conspicuous.

Delegating semantic interpretation to an external party means the verifier newly trusts the interpreting party: to accept the result, including vocabulary resolution, is to assume that that party performed them correctly---that is, to trust it. Such an introduction of trust should be made explicit under the formalization of trust (Section~\ref{sec:II_4}). Delegation lightens the burden but increases the number of parties to be trusted.

The role responsible for this layer is called the \emph{interpreter} in this model. It may or may not be the same party as the signature verifier; distinguishing the two lets us state clearly the difference between ``the signature can be verified but the meaning is not interpreted'' and ``the meaning too has been interpreted.''

\subsubsection{L3 Validation}
\label{sec:II_2_3}
Once semantic interpretation (L2) fixes what a claim asserts, whether the verifier accepts it is still undecided. Validation is the activity of evaluating an interpreted claim against the verifier's own criteria and deciding whether it may be accepted.

These criteria are called the \emph{validation policy}, determined by the verifier's purpose and context: for the same claim, one verifier may accept and another may not. Whereas the result of signature verification (L1) is constant independent of verifier and context, the result of validation depends on the policy the verifier adopts, reflecting its purpose and use; if the policy differs, so may the result.

Validation asks, among other things: is the issuer that emitted this claim acceptable as a party that emits claims of this kind; does the claim's content satisfy the verifier's conditions; and is the claim still in a state worthy of acceptance? Note that matters such as whether a signed object has been revoked or whether it is within its validity period are confirmations of objective facts, dependent on when validation is performed, and differ from the acceptance judgment: the facts can be confirmed mechanically, but whether one may accept on that basis is a separate question. Whether, having confirmed the certificate valid and not revoked, to impose a criterion such as ``accept only those issued within one month'' is set by the verifier's validation policy. Of these questions, whether an issuer is acceptable cannot be answered by the verifier alone: the materials for that judgment are supplied chiefly from the Constitution plane introduced in Section~\ref{sec:II_1} (details in Section~\ref{sec:II_3_1}). Which materials enter the acceptance judgment, however, is set by the validation policy, and the verifier may also refer to materials beyond the Constitution plane.

Here lies the contact point between the topmost layer of the establishment of verification (L3) and the Constitution plane on the acquisition side: validation can decide ``to which set does this issuer belong, and may that set be accepted'' by referring to the issuer set the Constitution plane defines.

The form of the acceptance judgment is not uniform: it may be a static match against a list the verifier holds in advance, or a trace of the trust chain linking the trust anchor to the issuer in question. These differences of form appear in the Constitution plane's design choice of how the set is constituted (Section~\ref{sec:II_3_1}). Common to both, validation refers to the constitution the Constitution plane defines to decide whether an issuer is acceptable.

That W3C VCDM 2.0~\cite{vcdm2} places validation outside its scope reflects validation's dependence on the verifier's intent. In the specification's normative terminology, verification is defined as confirmation that includes conformance to the specification, satisfaction of the securing mechanism, and, if present, a successful status check, while the means of validation---judging conformance to business requirements---are left out of scope. In this model, the confirmations included in verification decompose into results of L1 and L2 or factual inputs to L3, and validation corresponds to the acceptance decision of L3. What a specification can prescribe extends to confirming integrity and data origin (L1) and to the vocabulary framework on which L2 relies; what to accept is left to each verifier.

The role responsible for this layer is called the \emph{validator} in this model. The signature verifier that verifies signatures, the interpreter that interprets meaning, and the validator that decides validity are logically distinct roles: typically implemented in a single party, yet conceptually distinct activities that may be divided among different parties.

Once validation (L3) fixes acceptance, the establishment and acceptance of verification that this model handles is complete. Actually using an accepted claim---the consumption of a credential by a relying party---is an activity outside the model, premised on that acceptance: the model shows on what acceptance consumption rests, without entering into consumption itself. This acceptance presupposes that the verifier takes in the issuer set through its own policy (Section~\ref{sec:II_6}).

\subsection{Acquisition of Verification Materials}
\label{sec:II_3}
In this paper we call ``verification materials'' the collective term for what is supplied to the establishment of verification (L1--L3), and ``verification artifacts'' the individual fragments that compose those materials and that are treated as units of storage, retrieval, and bundling in acquisition.

\subsubsection{The Constitution Plane: Constitution and Establishment of a Community}
\label{sec:II_3_1}
When the verifier judges an issuer's acceptance in validation (L3), the verification materials are generally supplied directly from the issuer or from the community to which it belongs. The Constitution plane handles the arrangements a community shares and how, under them, the community is constituted and established. As stated in Section~\ref{sec:II_1}, the issuer set is one constituent element of a community, and the verification materials referred to in validation are constituted and managed under the community's agreement.

\paragraph{Constitution of a community's arrangements}
A community is constituted under the arrangements it shares. At their root lies the policy, through which governance works, fixing the governance of the community as a whole and of its constituent elements (how governance is exercised is beyond this paper's scope).

The policy includes the conditions for joining the set, the approach to managing the origin of trust, and the agreement on the protocols and data models used within the community. The constituent elements the policy prescribes are four---identifiers and namespace, the issuer set, the origin of trust, and other metadata---given either as independent elements or as consequences the policy yields.

\emph{Identifiers and namespace}: the identifiers that uniquely identify each party, including each issuer, within the set, and their namespace, or the definition thereof.

\emph{Issuer set}: the collection of issuers recognized in this community.

\emph{Origin of trust}: the set of propositions, shared in this community, assumed true as the starting point of the reasoning of verification. In this model, the key that serves as the handle for the reasoning of verification; in implementations it is represented as a trust anchor, a technical entity consisting of a signing key (or its digest) and associated data~\cite{rfc6024,rfc4034}.

\emph{Other metadata}: additional information shared as needed.

\paragraph{Derivation of the issuer set: logic and parameters}
Among these elements, the origin of trust and the issuer set are often conflated but differ in function: the origin of trust gives ``from where the reasoning of verification starts'' (a key), the issuer set ``who issues the credentials shared within that community.'' How the issuer set is identified is prescribed by the policy: in this model it is viewed as obtained by one step of derivation through the logic and parameters the policy defines. How concrete the logic and parameters are admits variation, of which enumeration is an extreme case.

\emph{Derivation}: feeding parameters, including context, into logic prepared in advance to judge whether a given issuer belongs to the issuer set. The judgment is not a simple match against a list but a dynamic resolution based on parameters and context. An example is OpenID Federation~\cite{oidfed}, where an entity's acceptance is resolved from information showing the chain and relations of declaring parties linking that entity to the origin; the logic is the chain-resolution procedure the federation shares.

\emph{Enumeration}: the case, within derivation, where the logic is simplified down to reference to a list and the judgment reduces to a static match. The parameter is an explicit list of issuers, against which the verifier judges membership in the set. The eIDAS Trusted List is one example.

The origin of trust, by contrast, is independent of this derivation: it is what the verifier accepts from the community and uses as the starting point of the reasoning of verification (Section~\ref{sec:II_4}, origin selection).

The choice of means for judging membership in the issuer set connects directly to the form in which validation (L3) is executed. Under derivation, the L3 acceptance judgment becomes the resolution of a chain, and the Constitution plane is expressed as the metadata and the relations among declaring parties composing that chain; under enumeration, it becomes a match against a list, and the information the Constitution plane provides is a static list.

How the policy defines the means of judging the issuer set is received differently on the issuing and accepting sides: the issuing side needs conformance to the policy to enter the issuer set; the accepting side, the verifier, needs acceptance of the policy. Merely receiving an issued credential does not complete validation (L3): only by also accepting that community's Constitution---the arrangements such as the issuer set and the origin of trust---can the verifier complete the processing across the three layers.

\paragraph{The word ``Constitution''}
As the name for this plane we use Constitution (constitution and establishment). What matters is less the static content of ``what is shared'' than the constitutional viewpoint of ``how the set is constituted and how it is established and governed.'' Constitution carries both senses---``the constituted state'' and ``the act of constituting''---and conveys the essence of this plane. One point needs care: a recursive structure is possible in which a community's own members are themselves further communities.

We deliberately avoid the term ``Trust Framework'': what it denotes is precisely what this model tries to analyze finely, and the term is used polysemously, without agreed definitions. The model does not deny the notion but refines it by decomposing it into multiple layers and planes.

\paragraph{The supply of verification artifacts concerning credential semantics}
Semantic interpretation (L2) interprets a claim according to the vocabulary and schema on which the credential relies; from where that vocabulary itself is supplied is an important viewpoint. A vocabulary such as CTDL is given not as inherent in each credential but as a framework, shared by the issuer set, for describing meaning.

The supply source of this vocabulary belongs to the Constitution plane. When the issuer set uses a particular vocabulary together, as ``the agreement on the data model used within the set'' (part of the policy), the vocabulary's definition becomes part of the information constituting the issuer set. From which issuer set's agreement the vocabulary derives, and from where it is fetched, is realized across the Constitution plane (the shared frame of meaning) and the Logistics plane (the delivery path of that definition). The framework includes artifacts of differing granularity or abstraction: the vocabulary itself (the language for description) and the definitions of individual credential types written in it (what is described in the language) can both become constituent information. A clue to whether a piece of information belongs to the Constitution plane is whether it is a type-level arrangement, fixed prior to an individual act of verification, and shared across the community.

Note that sharing a vocabulary (the Constitution plane), interpreting individual claims with it (L2), and judging acceptance based on membership in the issuer set (L3) are distinct activities: describing a credential's semantics finely and accepting its issuer as a member of the set, for instance, belong to independent axes.

\subsubsection{The Logistics Plane: The Delivery Path of Verification Materials}
\label{sec:II_3_2}
Signature verification (L1) completes in the verifying system's local computation if the full set of signed data from the trust anchor to the target credential is present---but that ``if present'' is a premise, not a given. The Logistics plane handles where each verification artifact, including the signed data, actually resides and by what means it reaches the verifying system.

\paragraph{The five elements of the Logistics plane}
The Logistics plane prescribes the means by which each verification artifact, including the signed data, reaches the verifying system. We capture this in five elements.

\emph{Element 1. Storage location}: the verification materials are not necessarily gathered in one place. The trust anchor is often pre-embedded (bundled) in the verifying system; the middle of the chain of trust may lie at issuer endpoints or third-party-operated repositories, with embedding into the credential and name-resolution systems such as DNS as further means; in the IHV model the target credential itself can be assumed to be in the holder's hands, and dynamically fetched information, such as revocation information, is placed at pre-designated locations. Where the artifacts composing the chain reside is the first condition determining how the verifying system discovers and retrieves them.

\emph{Element 2. Supply timing}: even for the same artifact, the nature of verification changes with when it reaches the verifying system. Distributed before verification (pre-provisioning), it need not be retrieved at verification time; handed over with the presentation of the credential (bundling at presentation), it is present at that point; retrieved only when verification is executed (runtime retrieval), verification depends on the success or failure of that retrieval---a viewpoint of availability becomes necessary at verification time. A configuration in which the means differ by artifact---the trust anchor pre-provisioned, the target credential bundled at presentation, revocation information retrieved at runtime---is entirely plausible.

\emph{Element 3. Supply protocol}: what communication procedure is used to retrieve each artifact. Retrieval over HTTP, retrieval involving name resolution, retrieval by dedicated issuance/presentation protocols---the means differ according to the artifact and its storage location.

\emph{Element 4. Use of standard protocols}: whether the protocol for supplying artifacts is specified as a standard or left to implementation. If standardized, verifying systems of different implementations can retrieve data by the same means, forming a basis for interoperability; without firm specification, each implementation uses its own means, and how artifacts arrive differs by verifying system.

\emph{Element 5. Retrieval path of dynamic information}: by what mechanism to retrieve information that can change over time, such as revocation information and various statuses. Whereas the preceding four elements chiefly target materials for static verification, this element targets time-varying information, and its method depends strongly on the nature of the target credential. Note that the five are not mutually orthogonal dimensions: element 4 is a property of element 3; element 5 is elements 1 through 3 applied anew to time-varying material; and elements 1 and 2 constrain each other (bundling, for instance, fixes storage location and supply timing at once). The intent of the enumeration is not independence as dimensions but coverage of the viewpoints for examining a supply design.

\paragraph{Orthogonality to the Constitution plane}
The Logistics plane is orthogonal to the Constitution plane: the Constitution plane prescribes ``what is shared'' (the constitution of the community), the Logistics plane ``how it reaches the verifying system.'' What the Constitution plane defines (the origin of trust, the issuer set, the vocabulary definition, and so on) is of no use unless it reaches the verifying system, and that delivery path is carried by the Logistics plane. Only when both planes are present do the materials needed for the establishment of verification gather at the verifying system.

\paragraph{Handling of dynamic information}
The retrieval of dynamic information, the fifth element of the Logistics plane, can differ in means by credential type: for a credential with no notion of revocation---a degree certificate that is not revoked once conferred, for instance---this element can be unnecessary, whereas for one where revocation, suspension, or renewal can occur at any time, such as a driver's license, revocation information is indispensable.

How dynamic information is used reflects requirements from the establishment side of verification: under what intent and conditions the signature over a claim was made (the payload and its meaning, handled by L1 and L2), and what the verifier requires in validation (the L3 validation policy), together prescribe what dynamic information becomes necessary. Managing the freshness of materials is not a task the Logistics plane completes alone; it is achieved only when each layer of the establishment of verification and the Logistics plane act in concert.

\paragraph{Network independence of verification}
Whether verification is independent of the network---making offline verification possible---is determined by the design choices of the Logistics plane. With a pre-provisioned trust anchor and intermediate artifacts bundled at presentation, for instance, offline verification holds. A design in which artifacts depend on runtime retrieval means that the verifying system depends, directly or indirectly, on network availability at verification time and on the legitimacy of the retrieval source, which affects the acceptance judgment.

\subsection{Formalization of Trust (The Shinken Framework)}
\label{sec:II_4}
The three layers carrying the establishment of verification (L1--L3) and the two planes carrying the acquisition of verification materials (Constitution and Logistics) do not operate independently: each of the five functions necessarily contains a moment of assuming-true a proposition that is not decidable by the verifying system from available computational evidence alone. The Shinken framework gathers these moments under a single viewpoint---not a function that a particular layer or plane carries, but a cross-cutting viewpoint acting on all five functions that makes visible on what trust each function of the model stands.

\subsubsection{Formalization of trust}
The verifying system is constituted as logic that derives a verification result from given inputs. It is required to settle its result deterministically (determinism) and, when a false positive or false negative occurs, to allow the cause to be identified (transparency).

The logic of verification, however, can contain propositions that are not decidable by the verifying system from the available computational evidence alone (hereafter undecidable propositions; the term is used in this operational sense, not that of computability theory). For instance, the proposition that ``the controller of some signing key states, as authentic, what it signs with that key'' concerns the operational reality of that party, and its truth cannot be confirmed by computation.

The difficulty is resolved by assuming the proposition in question true within the logic of verification; this introduction of an assumption is what the model calls trust. Fixing an undecidable proposition as an assumption lets the whole logic of verification be constituted as decidable logic. Nor is the introduction of trust limited to undecidable propositions: even for a decidable one whose criteria are complex and whose effect at revision is hard to explain, trust can be introduced and the criteria omitted, keeping the logic tractable. Trust is thus an unavoidable introduction of an assumption for making verification hold, and making its point of introduction explicit is itself a condition of transparency.

This structure admits a minimal notation. Let $E$ be the set of computational evidence available to the verifying system, $A$ the set of explicitly accepted assumptions, $R$ the set of inference rules, and $P$ the validation policy; the acceptance decision is then determined as $D(E, A, R, P)$. Determinism means that $D$ is uniquely determined for the same $E$, $A$, $R$, and $P$; transparency means that the origin of an outcome of $D$ can be traced separately to $E$ or to $A$. This notation serves as illustrative shorthand throughout the paper: it fixes a vocabulary for the inputs of the acceptance decision and for the separation between evidence and assumptions, and is not developed into a calculus in which derivations are carried out. The introduction of trust is precisely the operation of placing an undecidable proposition explicitly in $A$. An assumption placed in $A$ should, wherever possible, be recorded with its subject, scope, point in time, and responsible party: when an assumption breaks, this localizes the identification of the affected range and the re-examination.

\subsubsection{Ubiquity of trust across the functions}
The introduction of trust is not concentrated in a particular function but appears in each of the five functions.

In the three layers of the establishment of verification, the assumptions appear as follows. Signature verification (L1), to derive ``the controller of that key signed this payload'' from success, assumes the binding of a key to its controller and that the signature was applied under that controller's intent. Semantic interpretation (L2), when delegated externally, assumes that the interpreting party resolved the vocabulary correctly. Validation (L3) assumes that the materials the acceptance judgment relies on---the issuer set referred to and the trustworthy information obtained from multiple sources---hold as declared; the judgment deriving acceptance under a validation policy that deemed those materials worthy of trust likewise involves the introduction of an assumption.

The two planes of the acquisition of verification materials are no different: the Constitution plane assumes that the community is actually constituted and established under its arrangements (the policy) and that the issuer set and the origin of trust given from it are as the policy prescribes; the Logistics plane assumes that an artifact's retrieval source is legitimate and that the retrieved artifact satisfies the declared correspondence and freshness.

Each of these assumptions posits, as true, an operationally undecidable proposition. That trust gathers at no single point but is ubiquitous across all functions is itself the premise for discussing where to place roles.

\subsubsection{Origin selection}
Applying the formalization of trust to the trust anchor, the starting point of verification, yields one viewpoint on the structure of verification.

The verifying system's trust anchor is built into the logic of verification as ``the set of propositions assumed true'': not an objective fact given from outside, but the set of assumptions from which the verifier decided to begin its reasoning.

This re-frames, from origin selection, how the chain of trust is described. Naively, the chain runs from the issuer to the verifier. Seen from the verifier's side, however, the reasoning does not arrive from the issuer: it begins with the trust anchor placed inside the verifying system---a reference to the origin of trust---and is traced from there, through the issuer, to the target credential. The issuer-to-verifier description is the same chain captured with its origin taken on the verifier's side. All verification needs is that the full set of signed data from the origin of trust to the target credential be present; the party that raises that reasoning is the verifier.

\begin{figure}[tb]
\centering
\includegraphics[width=0.95\columnwidth]{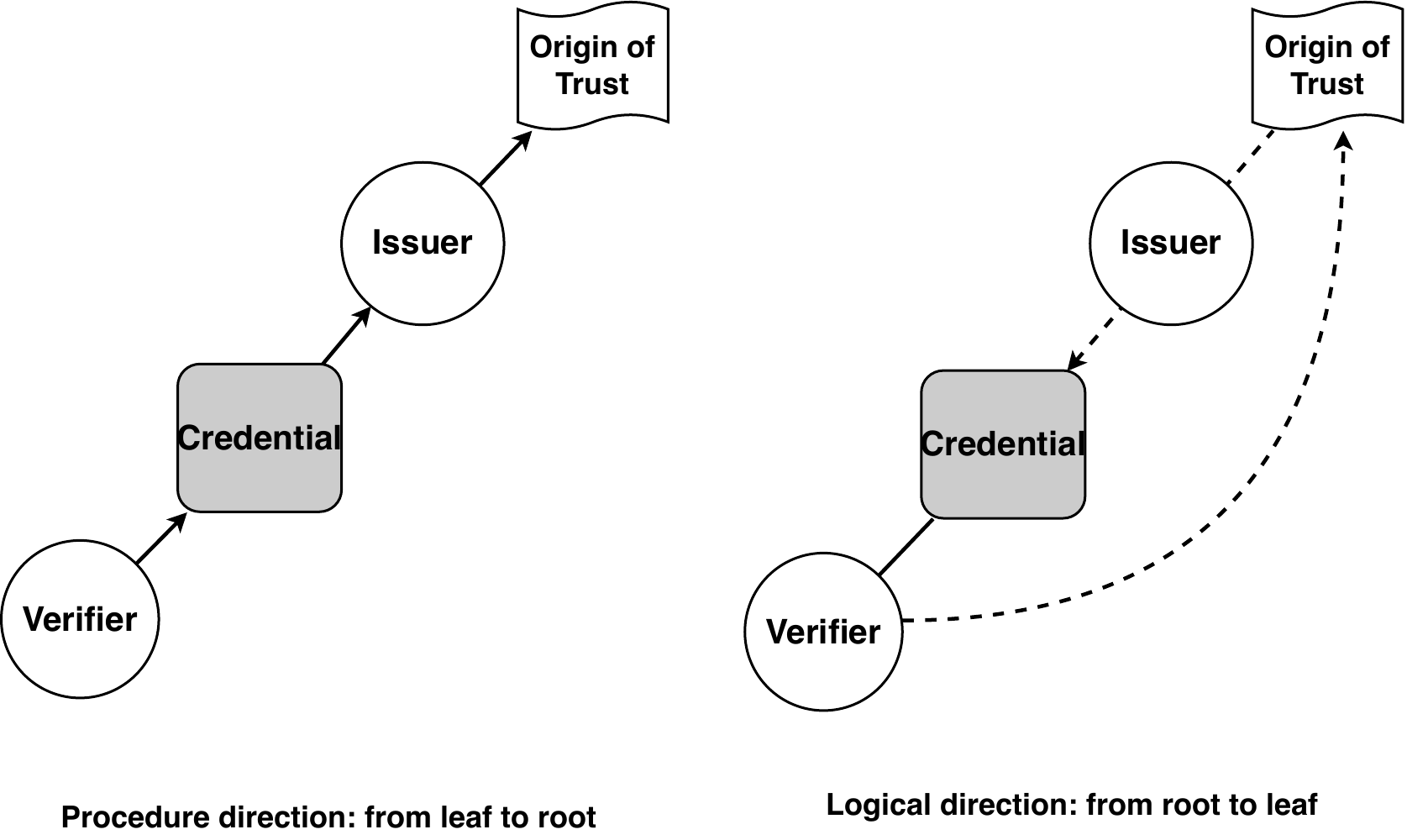}
\caption{Origin selection: the procedural direction (left: the verifier traces signing-key references from leaf to root) contrasted with the logical direction of inference (right: the verifier takes its trust anchor as the root and proceeds toward the credential it holds)}
\label{fig:origin}
\end{figure}

The logical direction of the reasoning must be distinguished here from the procedural direction of the implementation. Implementations often build a path by tracing signing-key references from leaf to root---from the target credential to its issuer and upward. Even so, the logical origin of the reasoning remains the root: the origin of trust the verifier assumed true. Which end a procedure builds from and where the logic places its origin are independent; origin selection shows the latter---it is the verifier's side that fixes the origin of the logic. This contrast is shown in Fig.~\ref{fig:origin}.

In IETF documents, a trust anchor is described as a technical entity: a signing key (or its digest) together with associated data. This paper uses the term according to the role that the entity plays in the verification logic. The technical entity is a representation of this set of propositions, and is also one of the components constituting the signed data that the Logistics plane handles.

\subsubsection{Mobility of the reasoning steps}
The viewpoint of origin selection gives a logical foothold for the question of where to place roles.

If verification is a reasoning chain originating at the verifier's trust anchor, then each step composing that reasoning---each of the five functions---may be executed by the verifier itself or delegated to a trusted other, with the result received. As long as the reasoning chain is composed on the verifier's side, who executes each step and where does not change whether the logic holds. That each function has freedom of placement (Section~\ref{sec:II_5}) derives from this mobility.

\subsubsection{Non-delegable functions}
One thing, however, cannot move.

It is the selection of assumptions. Which propositions to assume true---which trust anchor, whose declaration, to accept as the origin of the verification logic---fixes the origin of the reasoning chain itself, and only the verifier can do it. To entrust this externally is to hand over to another the decision of what to trust: the logical subject of the acceptance decision $D$ moves to the delegatee, and at that point it is no longer ``the verifier's own verification.'' The selection of assumptions cannot be delegated---a natural consequence of the verifier itself fixing the origin of the logic. Put differently, what cannot be delegated is not a computational operation such as signature verification but the responsibility for choosing which propositions to accept as assumptions.

Even the computation of signature verification is no exception to this mobility. With the signed data from the trust anchor to the target credential present, the computation can complete locally at the verifier (Section~\ref{sec:II_2_1}); it can also be entrusted to an external party with only the verification result received---a validating resolver in DNSSEC~\cite{rfc4033,rfc4034,rfc4035} is a typical example (in this configuration, the delegating side is required to receive the responses from the delegate over a secure channel). Yet delegation is not semantically neutral. If the delegatee returns evidence that can be independently re-verified, that evidence joins $E$ as computational evidence. If only an attestation of the result is returned, accepting the delegated result is to assume true that the delegatee performed the verification correctly: an addition of a proposition to $A$, simply a new introduction of trust. The computation can be delegated; the selection of the assumption that trusts the delegatee remains on the verifier's side.

The decomposition of roles (Section~\ref{sec:II_5}) targets the remaining functions, excluding this non-delegable one: of the model's functions, all but the selection of assumptions have freedom of placement. The next section discusses the dimensions that govern that placement.

\subsection{The Three Dimensions of Role Decomposition, and Availability}
\label{sec:II_5}
The preceding section showed that verification is a reasoning chain originating at the verifier's trust anchor, that each composing step has freedom of placement, and that the selection of assumptions cannot be delegated. For the remaining functions---the parts of each function's work that can be put outside or delegated---this section asks where, when, and how far they may be disclosed, in three dimensions. This freedom is not unconditional: changing the placement leads to the same acceptance decision only where the meaning of, and the dependencies among, the processing each function performs are preserved. The three dimensions pose separate questions but are not mathematically orthogonal coordinates: a single design choice may affect several dimensions at once.

\subsubsection{Dimension 1: Placement}
The first dimension is where to place a function in the information space: at the verifier itself, entrusted to an external proxy, shifted upstream, into the neighborhood of the issuer, or bundled into the credential and carried along. The word ``placement'' is used in two senses---a broad sense (freedom of placement) denoting the freedom that the three dimensions together define, and a narrow sense denoting this dimension, i.e., spatial placement. Alongside the other two dimensions, ``placement'' takes the narrow sense.

The object of placement is not limited to the two planes of the acquisition of verification materials (Constitution and Logistics); semantic interpretation (L2) is one as well. Resolving a vocabulary need not always be completed at the verifier's edge: it can be delegated externally, or only the parts needed referred to. The ``heaviness in processing'' of a highly expressive vocabulary is thus not inherent in the vocabulary itself but governed by where the work of interpretation is placed: performed at the verifier's edge, the burden there grows; placed elsewhere, the verifier's burden changes.

Where to place a function can be organized from the viewpoint of three asymmetries.

\emph{Resource asymmetry}: resolving a vocabulary or verifying a chain requires commensurate computational resources and code. Whether a party can meet this need varies, and can prompt delegation to a party that can secure the resources.

\emph{Trust asymmetry}: each time a function is delegated, the verifier newly trusts the delegatee. Delegation lowers the verifier's burden while increasing the parties it must trust. If many verifiers delegate to the same proxy, a reconcentration of trust can arise there.

\emph{Information asymmetry}: whether the information a function needs can be fixed in advance or only at runtime changes the placement. A function depending on information fixable in advance can be shifted upstream, toward the issuer side. A function depending on runtime state cannot be served by prior placement alone, but this does not mean the processing must sit at the verifier's hands: it can also be delegated to an external service reachable at runtime.

\subsubsection{Dimension 2: Timing}
The second dimension is when to begin the reasoning. It may be raised while a session with the counterpart is ongoing (the authentication flow), or a signature may be applied at issuance and verified at presentation (credential issuance and presentation): the same tracing of a chain of trust is done within a session in the one case and split between issuance and presentation in the other. From the viewpoint of origin selection, the reasoning on either side is isomorphic; only the timing of raising it differs. This dimension concerns when to raise the reasoning---independent of when the verification materials are supplied (retrieved at runtime or bundled at presentation), the supply-timing choice of the Logistics plane. Even a design that raises the reasoning at presentation may retrieve materials at runtime or bundle them at presentation.

\subsubsection{Dimension 3: Disclosure}
The third dimension is whether the verification materials may be disclosed: whether the materials needed for verification (the list of issuers and the origin of trust held by the Constitution plane, the storage locations and supply paths prescribed by the Logistics plane, and so on) may be referable by anyone or should go only to limited parties differs by ecosystem and context.

This dimension has two poles. At one, publicity of the verification materials is itself a requirement: in uses aimed at letting the public confirm the origin of emitted information, the issuer set and its constitution may---indeed should---be referable by anyone. At the other, disclosing the materials is not permitted in itself: where transaction or supply-chain relations appear in them, the very existence of those relations is information to be concealed, and disclosure would expose the structure independently of verification; verification is needed, yet the materials must be disclosed selectively, only to legitimate parties. For the same mechanism of verification, the required design differs entirely between the two. Even when the content of the materials is kept confidential, runtime retrieval can expose metadata---the retrieval endpoint, the time, the frequency---to observation. The publicity dimension covers not only the disclosure of content but also this range of observation.

\subsubsection{Consolidating the availability discussion into the Logistics plane}
In verification, ``the availability of the retrieval source of an artifact at verification time'' is often spoken of as an independent property. In this model, however, that verification-time dependence is organized as primarily shaped by the choice of supply timing in the Logistics plane.

Suppose validation takes the form of deriving the issuer set. If the artifacts the derivation needs are, by design, retrieved at verification time, the establishment of verification depends on the availability of the retrieval source. If enough artifacts to make the derivation hold are instead bundled into the credential at presentation, the verifier can derive at runtime without querying anyone, removing the runtime dependence on availability (the dependence does not vanish; it shifts to the burden at issuance, distribution, and update time). What decides it is not the choice of the form of derivation but the timing choice in the Logistics plane---whether the artifacts are retrieved at runtime or bundled at presentation.

The discussion also bears on privacy. A design that fetches materials at runtime discloses to the retrieval source ``who is now verifying whose what''; this can be concealed by not retrieving---that is, by bundling into the credential. The design choice between depending on runtime retrieval of artifacts and completing with bundling thus decides the dependence on the availability of the retrieval source, and also how far the verifier exposes its own verification context externally.

\subsection{The Whole Model in Overview, and the Position of the ``Registry''}
\label{sec:II_6}
Up to here, the model has been built on two decompositions. The first, the decomposition of function, divides the activity surrounding verification into five functions: three layers (signature verification, semantic interpretation, validation) and two planes (Constitution, Logistics). The second, the decomposition of placement, captures the parts of those functions, excluding the non-delegable function, as a freedom of placement along three dimensions (where to place spatially, when to raise the reasoning, how far to disclose). The former asks ``what work it is,'' the latter ``where, when, and how far it may be disclosed.'' This section surveys the model on that basis.

\subsubsection{Conditions for the establishment of verification}
Verification holds only when two planes of differing nature are present together: the Constitution plane, which prescribes how a community is constituted and established, and the Logistics plane, which delivers to the verifying system both the content of that constitution and the credential itself. Through them, the verifier obtains the trust anchor to be built into the verifying system and the chain of signed data reaching the target credential. The two planes are a division of function, not a requirement for two independent organizations or separate protocols: the same party or the same implementation may carry the functions of both planes.

The constituent elements of the two planes differ in freedom of placement: each function, excluding the non-delegable selection of assumptions, can be placed according to the three dimensions of role decomposition. In this model, verifiability can be viewed as the relation of these two planes and as a freedom of placement.

\subsubsection{The relation of the ``registry'' to the discussion of this model}
Around the supply of verification materials, the word ``registry'' is often used. We show what this word can denote in the model.

A registry is not the Constitution plane itself: it is one implementation form of that plane in the enumeration form---a data store that, as an authority, holds and declares the issuer set. Mechanisms providing an explicit enumeration of a list of issuers correspond to it. The Trust over IP Trust Registry Query Protocol (TRQP)~\cite{trqp} standardizes queries against implementations of this enumeration form, and is consistent with this model's organization.

By contrast, the collection of declarations obtained under the derivation form is not appropriately called a ``registry.'' The word carries connotations of a single authority, a centralized ledger, and a static reference source---connotations that do not fit a structure deriving membership by tracing distributed declarations, as in a federation. The Constitution plane can, as a concept, take both forms; a registry merely expresses the enumeration form. In practice, however, structures that derive membership by tracing distributed declarations, as in a federation, are sometimes also called registries. When the word is used, it should at least be made explicit which of the following is meant: (a) a static enumeration, (b) relations and declarations serving as inputs to derivation, (c) a repository or API for delivery, or (d) a combination of these. This decomposition avoids inferring issuer-acceptance rules or interoperability from the single word ``there is a registry.''

\subsubsection{Independence of declaration and acceptance}
On the other hand, a party's declaring the issuer set and the verifier's taking that constitution into its own verification logic are independent.

Acceptance is not an implicit, automatic process; it is mediated by the verifier's validation policy. Whether to trust some party's declaration, or to accept some issuer set, is set as policy, and the verifier decides acceptance accordingly. That a registry declared does not, in itself, mean the verifier accepts. This independence also appears explicitly on the specification side: DIF's Credential Trust Establishment states that publishing the governance file itself places no requirement on other participants to accept it~\cite{difcte}.

This separation corresponds to two elements of the model. In validation (L3), a registry's declaration is an input to the acceptance judgment, not the judgment itself; a validation policy may have multiple inputs, referring to multiple trustworthy materials from different sources and integrating them into one acceptance judgment. And in the formalization of trust (Section~\ref{sec:II_4}), to trust some party's declaration is itself the introduction of trust into an undecidable proposition---an assumption introduced explicitly as policy rather than given implicitly---and the choice of whose declaration to trust is the non-delegable selection of assumptions.

\subsubsection{A registry does not carry the delivery path}
A registry is an authority that holds and declares what is correct; it does not carry how that content reaches the verifying system. Artifacts are carried by the Logistics plane---the registry does not transport. The model's picture that the Constitution plane (with a registry as its enumeration form) and the Logistics plane are orthogonal rests on this distinction. A registry is the substance of a declaration placed on the side of the issuer set; the verifier refers to and accepts it through its own policy, the acceptance is fixed in validation (L3), and beyond it a relying party consumes the credential (Section~\ref{sec:II_2_3}). Declaration, acceptance, and consumption are separate activities of the declaring party, the verifier, and the relying party, respectively.

\subsection{Consequences of Being Verifiable, and a Worked Example}
\label{sec:II_6_conseq}
We present seven consequences derived from the conditions of being verifiable. The seven are not claims of equal strength; they divide into three kinds: definitional corollaries that follow directly from the definitions (the first and fourth), operational implications that follow from design conditions (the second, third, and fifth), and design trade-offs whose evaluation varies with placement (the sixth and seventh).

First (a definitional corollary), the scope within which verifiability holds is bounded by the set of arrangements and trust assumptions the verifier accepts: a chain from the trust anchor to the target credential can hold only where the verifier accepts the constitution making it hold---that is, within the community. Following directly from the definition of community, its significance is to carve out cross-boundary connection as an explicit object of analysis. In this sense, verifiability is not an absolute property: it is relative to the verifier, the use, the point in time, and the accepted arrangements and trust assumptions---the same credential can be verifiable to one verifier and unverifiable to another. Nor does this boundary always coincide with a legal or geographic enclosure; it is drawn by the acceptance of arrangements and assumptions.

Second (an operational implication), verification is reasoning that starts at the verifying system's trust anchor and proceeds toward the target credential (Section~\ref{sec:II_4}, origin selection).

Third (an operational implication), once the trust anchor and the chain of signed data are present at the verifying system, verification completes as a local computation and requires, in principle, no runtime inquiry to any external party.

Fourth (a definitional corollary), verifiability holds only when the two planes, Constitution and Logistics, are both designed; neither alone suffices. The two planes are orthogonal yet not unrelated: under the derivation form, the chain of declarations deriving membership composes the content of the Constitution plane, yet if the artifacts making that chain hold are bundled at presentation, even they can be carried by the Logistics plane. The timing choice of the Logistics plane thus governs the form in which the content of the Constitution plane reaches the verifying system.

Fifth (an operational implication), offline verifiability is determined by the choices of storage location and supply timing in the Logistics plane (Section~\ref{sec:II_3_2}, network-independence of verification). If, however, the validation policy demands an up-to-date status check at execution time, bundling the materials alone cannot make verification offline: offline verifiability holds when both the Logistics choices and the policy's freshness requirements permit it.

Sixth (a design trade-off), how the ``weight'' of a given placement is evaluated is relative to how the verifier's role is set: the ``weight'' of placing vocabulary resolution at the verifier's edge grows under a role setting that performs it there; change the placement, and the evaluation changes too.

Seventh (a design trade-off), the model captures two important sources of privacy risk, and their interaction. One is leakage of the verification context: a design that retrieves materials at runtime exposes the verification context to the retrieval source---a matter of supply timing in the Logistics plane. The other is the disclosability of the verification materials themselves: whether they may be public or require limited disclosure is shown by the disclosure dimension (Section~\ref{sec:II_5}). Where disclosing the materials is not permitted in itself, runtime retrieval compounds the exposure, and leakage becomes most severe; between uses where disclosure is permitted and where it must not be, the designs required for privacy are opposite.

The relation of these two planes, the placement along the three dimensions, and the consequences drawn from them organize, under one model, the phenomena presented at the outset (the four observations of Section~\ref{sec:intro}, and the many senses in which terms such as registry are used).

\subsubsection{Worked example: tracing a single diploma}
\label{sec:II_6_walk}
We now trace the decomposition end to end through one concrete example: a university (issuer) issues a diploma to a graduate (holder), who presents it to a company (verifier) in a hiring process.

In signature verification (L1), the verifying system follows the reference to the verification key bundled with the diploma and verifies the signature. What is confirmed is that the payload has not been altered since signing and that the controller of the corresponding signing key signed it---a confirmation that yields the same result no matter who performs it.

In semantic interpretation (L2), the payload is interpreted against the vocabulary it relies on: which institution awarded which degree, to whom, and under which criteria. The vocabulary is not intrinsic to each individual diploma; it is supplied as something the community shares.

In validation (L3), acceptance is decided against the verifier's validation policy---for example, ``accept only what an accredited institution of higher education has issued.'' The materials for judging whether the issuer belongs to that set cannot be prepared by the verifier alone; they are supplied from the Constitution plane.

Suppose the Constitution plane in this example takes the enumeration form: a list of accredited institutions (a trust list) is maintained under the policy and shared together with the identifier namespace and the origin of trust. The verifier judges the issuer's membership by matching against this list.

On the Logistics plane, the origin of trust is provisioned to the verifying system in advance, the diploma and the intermediate signed data are bundled at presentation, and revocation information is unnecessary (we take a degree to be of the type that, once awarded, is not revoked). With these choices, verification completes at the point of presentation without depending on the network.

Retracing this sequence through the Shinken framework brings out the propositions posited as true at each step: the binding of key and controller (L1), the correct resolution of the vocabulary (L2), the list's declaration being as the policy prescribes (Constitution), and the bundled artifacts satisfying the declared correspondences (Logistics). The choice of which of these to accept as assumptions remains on the verifier's side.

The same example takes a different shape as the choices along the three dimensions change: delegating vocabulary resolution to an external interpreting agent changes the placement and adds a new introduction of trust to be made explicit; a credential type with revocation brings in runtime retrieval, and supply timing begins to bear on availability and privacy; a list requiring restricted disclosure lets the disclosure dimension reshape the whole design. The model's vocabulary describes these as differences of choice over the same skeleton.

\subsection{Generalization to Combinations of Multiple Credentials}
\label{sec:II_7}
The discussion so far has been built around a single credential (Section~\ref{sec:II_1}): the three layers and two planes form a skeleton stating by what logic a single credential is verified. Real verification, however, often involves multiple credentials---one might even say the situation truly to be discussed is one where a credential confirming identity and one asserting an attribute (a national ID and a learner credential, say) satisfy the verifier's purpose only in combination. This section generalizes the model to such combinations.

The generalization requires no new layer; it can be expressed as an extension of the structure already described, in particular the integration of multiple sources in validation (Section~\ref{sec:II_6}).

\subsubsection{Multiplexing of the Constitution/Logistics pair}
First, the Constitution and Logistics planes are not one pair for the whole model: a corresponding pair exists per credential, each with its own logic and parameters for deriving the issuer set (via the Constitution) and its own delivery path (Logistics). Combining credentials from different governance systems, such as a national ID and a learner credential, hands multiple Constitution/Logistics pairs to the verifier. This does not mean, however, that each pair is a distinct entity: the same Constitution may be shared by multiple credentials. Even then, the picture is unchanged---for each credential, the Constitution and Logistics on which it relies are identified.

The acquisition of verification materials in combined verification is therefore assembly from multiple supply systems, not retrieval from a single one: the verifier gathers, under its own validation policy, materials arriving by different paths, with different constitutional declarations and trust anchors per credential. The picture of the two planes is preserved per credential while the whole is multiplexed.

\subsubsection{The chaining of the establishment of verification: the basic form}
\begin{figure}[tb]
\centering
\includegraphics[width=0.95\columnwidth]{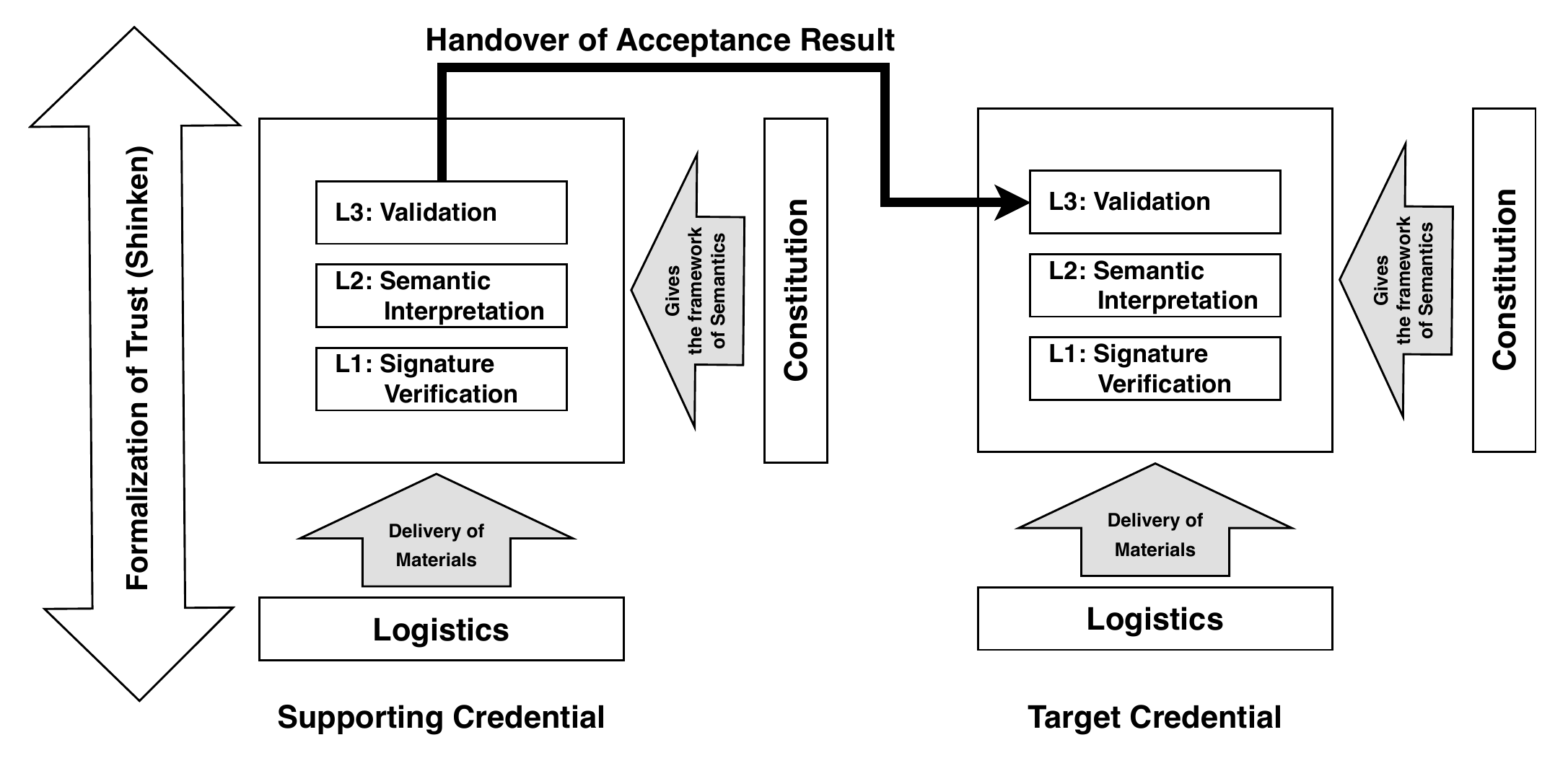}
\caption{An example of nesting in the establishment of verification under combined verification. A three-layer stack with its Constitution/Logistics pair is placed once each for the supporting credential and the target credential. The result of the supporting credential's validation (L3) is passed as an input to the target credential's validation. Each stack is given its interpretive frame under its own Constitution and follows an independent set of arrangements.}
\label{fig:composition}
\end{figure}

How are the verifications of multiple credentials related? In the basic form, the establishment of verification of one credential becomes an input to the validation of another: one completes its own three layers (signature verification, semantic interpretation, validation), and the other's validation refers to that fixed result as one of its verification materials. Typically, the verification of a credential confirming identity holds first, and, premised on that, the acceptance of an attribute credential is judged. This basic form is shown in Fig.~\ref{fig:composition}.

This form simply generalizes the picture of Section~\ref{sec:II_6} (integrating multiple trustworthy materials into one acceptance judgment). ``Materials'' has so far meant those directly involved in verification, such as declarations and revocation information; the generalization is that the result of the establishment of verification of another credential---complete with its own three layers, two planes, and Shinken framework---can itself be such a material. No new layer is stacked on top of validation; the input to validation can simply be the output of another establishment of verification.

Origin selection (Section~\ref{sec:II_4}) is preserved under this generalization too. The verifier's reasoning is now not a single chain rooted at one trust anchor but a structure in which multiple chains, each rooted at a different trust anchor per credential, merge at the verifier's validation. The roots multiply, but every chain still starts from the assumptions of trust the verifier itself holds: the origin of the reasoning remains on the verifier's side even with multiple roots.

\subsubsection{A special form, and making the introduction of trust explicit}
Against this, there can also be a form in which multiple credentials each complete signature verification and semantic interpretation, and a single validation accepts the whole at once---a special case of the basic form. Instead of fixing each credential's acceptance judgment independently, one validation policy treats all credentials as inputs of equal standing and decides acceptance in a single judgment. Only the number of judgment stages differs; gathering materials from multiple supply systems and consolidating them under the verifier's policy is shared with the basic form.

In either form, the objects the Shinken framework (Section~\ref{sec:II_4}) requires to be made explicit increase. Accepting another credential's verification result as an input to validation is to assume true that that verification was performed correctly---a new introduction of trust. The assumption tends to be placed tacitly when the referred verification was performed by a different party, that is, when only the result of a delegated verification is received; the merge point of combined verification is thus a typical position where Shinken requires trust to be made visible. And where the verifying party's consolidated acceptance itself feeds yet another party's validation, a need can arise to make explicit, to that party, which verification results were consolidated under which assumptions.

This concludes the presentation of the model. The next section applies it to the stacks presented earlier and attempts an analysis.

\section{Application and Evaluation}
\label{sec:application}
We apply the model to the four kinds of stacks observed in the learner-credential domain~\cite{edcredi}, interpret the phenomena presented in the introduction, and then cross-compare existing ecosystems, including authentication systems, along the three dimensions of role decomposition. The learner domain serves as the main concrete example; the model itself is not limited to learner credentials. This section's evaluation is an analytical demonstration based on public documents; it does not rank the stacks' performance or maturity. The four stacks are not at a uniform level of abstraction or completeness, ranging from base specifications alone to an ecosystem equipped with regulation, a trust model, and protocol profiles. The purpose of this section is not to rank them but to make visible, in a single vocabulary, which functions each stack specifies itself and what it leaves to external parties in composing a complete verification stack. A conceptual model cannot be evaluated through quantitative performance measurement; this section evaluates against three criteria: whether it can explain the phenomena of Section~\ref{sec:intro} (explanatory power), distinguish design differences across stacks in a single vocabulary (discriminating power), and derive design implications from the analysis (derivation of implications). Such criteria-based evaluation follows the notion of evaluation in design science~\cite{hevner} and taxonomy development~\cite{taxonomy}.

The judgment criteria of this section's tables are as follows. In Table~\ref{tab:stackeval}, ``explicit'' means that the constituent element is specified in normative statements of a public specification or regulatory document; ``undefined,'' ``out of scope,'' and ``not specified'' mean that no specification could be confirmed in the stack's public documents. All judgments rest on the analyses in Sections~\ref{sec:III_1} to \ref{sec:III_4} and the specifications cited there (reference time points: Section~\ref{sec:III_8}); the judgments in Table~\ref{tab:related} likewise rest on the explicit scope of description of the works concerned.

\subsection{Stack A: Open Badges / CLR (1EdTech)}
\label{sec:III_1}
Constitution plane: among the community's constituent elements, the sharing mechanisms for the ``origin of trust'' and the ``issuer set'' are undefined; policy too is left to each issuer, and no community-wide policy exists as a structure.

Logistics plane: it relies on HTTP reference to issuer URLs, with weak specification of storage location and supply path at the protocol level.

As a result, the verifying system comes to depend implicitly on the Web PKI TLS certificate chain---which shows ``that this domain communicates legitimately,'' not ``that it is legitimate as an issuer of learner credentials.'' The analysis targets the operational reality of the Open Badges 2.0 generation. In the current-specification generation, Open Badges 3.0~\cite{ob3} and the Comprehensive Learner Record Standard 2.0~\cite{clr2} prescribe VCDM-conformant signature representations, and CLR 2.0 reuses the EndorsementCredential---a credential carrying third-party endorsement---defined by OB 3.0. The credential-transfer API was historically specified as the Badge Connect API of OB 2.1; in OB 3.0 the Open Badges API, integrated into the specification body, provides the standard path. In addition, the TrustEd Credential Coalition (announced in March 2024 as the TrustEd Microcredential Framework and later renamed) provides a framework that supplements metadata quality outside the specification layer~\cite{trustedcred}. Even so, use-specific issuer authorization, the trust anchors to adopt, and the final validation policy are not fixed by the specifications, and the Constitution plane's sharing mechanism remains undefined at the specification layer---a gap that marks the boundary of responsibility between the specification layer and the ecosystem layer rather than a defect. The breakdown of email-address-based learner identification in Open Badges 2.0 is due to this deficiency of the two planes, in addition to an email address not proving control by a particular subject (SP 800-63B~\cite{sp80063b}).

\subsection{Stack B: Unprofiled DID/VC (W3C)}
\label{sec:III_2}
Constitution plane: what the VCDM and DID~\cite{did1} specifications provide extends to the structure of signed data and the abstract definition of identifiers; the sharing mechanisms for ``policy,'' ``origin of trust,'' and ``issuer set'' are outside the scope of the W3C specifications and left to the application side.

Logistics plane: it is DID-Method-dependent, with resolver-based runtime retrieval as the representative form (a DID Document can also be obtained through peer methods or prior distribution). The proliferation of more than 260 DID Methods~\cite{didmethods} can be described as the consequence of no convergence pressure being applied to the Logistics plane.

\subsection{Stack C: EDC/ELM/EUDI-Wallet (European Commission)}
\label{sec:III_3}
Constitution plane: the trust lists of eIDAS 2.0 provide the constituent elements in a nearly complete form. Policy is prescribed as the eIDAS regulation; identifiers and namespace as per-member-state namespaces. The issuer set is enumerated explicitly---in Trusted Lists for QEAA and PuB-EAA providers, and in a List of Trusted Entities (LoTE) for PID Providers. The origin of trust is the trust anchors these lists carry, discoverable through a common trust infrastructure maintained by the European Commission.\footnote{The specification analysis of Stack C is based on ARF v2.9.0~\cite{eudiarfv2}, the current version as of the literature-access date (July 11, 2026). Although this postdates the observation window of the operational analysis (November 30, 2025; Section~\ref{sec:III_8}), the revisions since v2.0 chiefly refine the trust-list structure and systematize the catalogues, and do not change the conclusions of this section.} Providers of non-qualified EAAs are also subject to registration by a Member State Registrar, but they are not notified to the European Commission, and their trust anchors are not automatically included in a Trusted List or the LoTE. The ARF recommends (SHOULD) that the applicable Attestation Rulebook define a domain-specific mechanism for obtaining trust anchors, and methods other than the LoTE are permitted.

Logistics plane: OID4VCI~\cite{oid4vci}/OID4VP~\cite{oid4vp} prescribe the issuance and presentation protocols of credentials, the Status List prescribes the retrieval path of dynamic information, and a Trusted List distribution protocol prescribes the supply of the Constitution plane's content. OID4VCI itself leaves out of scope how trust in the signer of signed issuer metadata is established (its Section 12.2.3); in the EUDI ecosystem this is supplemented by the ARF, which prescribes certificate chains leading to the trust anchors of Access Certificate Authorities~\cite{eudiarfv2}.

Within the scope treated in this paper, this is the only stack in which both planes are designed. Offline verifiability also holds, within the profiles analyzed here, by combining the holder-bundled type of mdoc/SD-JWT VC with the pre-provisioned type of the Trusted List---under configurations where status information, including revocation status and the currency of the pre-provisioned list, is either not required or is available under the freshness policy the deployment assumes. Its community boundary, however, is fixed within the EU; connection beyond it requires additional inter-community mapping and mutual acceptance arrangements.

From the model's viewpoint, the design of Stack C is an example other stacks should take as reference, though the handling of the community boundary remains a topic for further development.

\subsection{Stack D: Platform-proprietary}
\label{sec:III_4}
Constitution plane: a single platform operator prescribes the constituent elements.

Logistics plane: it completes within the platform's own internal protocol.

This is a vertical integration in which a single operator holds both planes. It functions as a community, but its governance is concentrated in a single party, and interoperability beyond the boundary depends on the platform operator's intent.

Table~\ref{tab:stackeval} summarizes the analyses of Stacks A--D in a unified format using the model's vocabulary: aligning the Constitution plane's constituent elements (policy, issuer set, origin of trust) and the Logistics plane's specification status under the same viewpoints makes the design differences across the stacks readable within a single vocabulary.

\begin{table*}[!t]
\caption{Unified assessment of the four learner-credential stacks: specification status of the Constitution and Logistics planes}
\label{tab:stackeval}
\centering
\footnotesize
\begin{tabular}{@{}p{2.0cm}p{3.5cm}p{3.5cm}p{3.5cm}p{3.5cm}@{}}
\hline\hline
\textbf{Viewpoint} & \textbf{Stack A: Open Badges/CLR} & \textbf{Stack B: unprofiled DID/VC} & \textbf{Stack C: EDC/ELM/EUDI-Wallet} & \textbf{Stack D: platform-proprietary} \\ \hline
Policy & Left to each issuer; no community-wide policy & Outside W3C specification scope (left to the application) & Explicit, as the eIDAS regulation & Prescribed solely by the operator \\
Issuer set & Sharing mechanism undefined & Out of scope & Explicit: Trusted Lists (QEAA/PuB-EAA); a LoTE (PID Providers) & Prescribed solely by the operator \\
Origin of trust & Undefined (implicit dependence on the Web PKI TLS chain) & Out of scope & Explicit: trust anchors in the Trusted Lists/LoTEs (discovered via the common trust infrastructure) & Prescribed solely by the operator \\
Delivery path (Logistics) & HTTP reference to issuer URLs; storage and supply weakly specified & DID-method-dependent; runtime retrieval via resolvers as the representative form & Prescribed by OID4VCI/OID4VP and a Trusted List distribution protocol & Within the platform's internal protocol \\
Retrieval path of dynamic information & Not specified & Not specified & Status List & Within the internal protocol \\
Design status of the two planes & Neither is sufficient & Neither is sufficient & Both planes designed (the only case within the scope of this paper) & A single operator holds both planes \\
\hline
\end{tabular}
\end{table*}
\subsection{Worked Example: Sharing the Same Data Model Is Not Sufficient for Interoperability}
\label{sec:III_example}
How the coexistence of stacks appears as a failure of interoperability can be confirmed with one concrete example. An OpenBadgeCredential of Open Badges 3.0 is a verifiable credential conforming to the VCDM~\cite{ob3}. Suppose it is presented to a verifier---corresponding to Stack B---that implements only the base specifications of the VCDM and DID Core. At L1, the signature cannot be processed unless a compatible proof scheme and a way to obtain the verification key have been selected. At L2, without an arrangement for interpreting the OB 3.0 vocabulary (Achievement and the like), the meaning of the claims is not determined. At L3, the materials for judging which issuers may be accepted (the issuer set and the origin of trust) are not supplied by any Constitution plane.

That is, this failure is positioned not as a mismatch on the single item ``whether the same JSON-LD or VC is used,'' but as the absence, somewhere in L1--L3, Constitution, or Logistics, of a shared choice, material, arrangement, or policy. The agreement ``the same data model'' that element-by-element comparison exhibits shares only part of the premises of L2.

\subsection{Cross-Ecosystem Comparison by the Three Dimensions of Role Decomposition}
\label{sec:III_5}
Stacks A--D are cases confined to the learner-credential domain. This section uses the three dimensions of role decomposition (placement, timing, disclosure) to cross-compare existing ecosystems, including authentication systems; widening the comparison is warranted because the discriminating power of the three dimensions appears most clearly among ecosystems whose choices of timing and disclosure diverge. Stack C (Section~\ref{sec:III_3}) corresponds to the EUDI / EUDI-Wallet row and is analyzed both from the specification status of the two planes and from its position on the three dimensions. We also include Originator Profile (OP)~\cite{opf}, a proposal for originator-attribute disclosure and content protection in media publishing, although its use case differs from credential issuance; OP is a system of static attribute disclosure, premised on the information being public. The authors are involved in the design of OP; we state this explicitly and include it in the comparison. Table~\ref{tab:cmp} presents this cross-ecosystem comparison.

\begin{table*}[tb]
\caption{Cross-ecosystem comparison along the three dimensions of role decomposition}
\label{tab:cmp}
\centering
\footnotesize
\renewcommand{\arraystretch}{1.2}
\begin{tabular}{p{0.13\textwidth} p{0.12\textwidth} p{0.16\textwidth} p{0.42\textwidth}}
\hline\hline
\textbf{Ecosystem} & \textbf{Timing} & \textbf{Disclosure} & \textbf{Placement characteristics} \\ \hline
eduGAIN (SAML/MDS) & session time & public (aggregated distribution) & A pre-aggregator aggregates and signs the distribution; the verifier accepts the aggregated set en bloc. Simple but centrally aggregated: pre-distribution removes dependence on a retrieval source at verification time, at the cost of aggregation freshness and trust concentrated in the aggregator. \\
OpenID Federation (per-entity) & session time\newline (runtime resolution)\newline or presentation\newline (bundled) & public (resolvable) & The verifier resolves dynamically or delegates to a resolver; alternatively the entity supplies its own trust chain with the presentation, which removes runtime retrieval at the cost of freshness bounded by the chain's expiry. Distributed. Runtime resolution incurs a fetch/privacy cost and depends on the availability of the retrieval source. \\
EUDI / EUDI-Wallet & issuance + presentation & public (Trusted List) & Three functions separated: Trusted Lists (eIDAS) as the issuer trust layer, VCI as issuance, VP as presentation. The trust chain is verified at presentation; bundling there removes dependence on the issuer's runtime availability. \\
Originator Profile (reference) & issuance + viewing & public availability is a requirement & The verifier (a browser extension) is kept minimal; materials shift to the issuer side or are bundled with the content. Bundled verification is self-contained, requiring no reachability to the issuer at verification time; dynamic retrieval from the verifier side is also supported. \\
\hline
\end{tabular}
\end{table*}
The same trust-establishment mechanism, Federation, appears in both the authentication context (resolving the counterpart) and the credential context (resolving the signer); that the same word appears in two contexts is one cause of the terminological confusion arising across multiple ecosystems.

As a function set of the three layers, the two planes, and the Shinken framework, these ecosystems do isomorphic work but differ in the choices along placement, timing, and disclosure; the model can provide a vocabulary that describes functional isomorphism and placement difference at the same time. The same organization applies to the data-space architecture cited earlier~\cite{vaia}, which lets three choices of discovery mechanism coexist---derivation by runtime resolution, self-hosted presentation of metadata, and bundling with the content---and is thus an instance of a Logistics-plane design that accommodates several choices at once.

\subsection{Demonstrating the Model's Interpretive Power}
\label{sec:III_6}
Two examples show that the model has the power to interpret existing systems and discussions: eduGAIN~\cite{edugain}, where the three dimensions of role decomposition analyze how the same function appears at different timings and placements; and the Registry of the CTDL ecosystem, where decomposition into the two planes analyzes one word bundling entities of different planes.

\subsubsection{eduGAIN: distinguishing authentication and credential}
eduGAIN was originally a federation of authentication/authorization (an IdP passes in-session attributes to an SP as SAML assertions), and it operates as an interfederation service in which the MDS aggregates, validates, signs, and republishes the participating federations' metadata~\cite{edugain}; those attributes are not portable credentials. Learner records such as a diploma, by contrast, are credentials ``issued, held by the person, and presented at an arbitrary time''; the two are different activities.

eduGAIN has been running a pilot of OpenID Federation since July 2025~\cite{edugainpoc}; this is a trial toward updating the trust-establishment mechanism of authentication and is not directly linked to credential issuance. Yet because Federation appears as a ``trust layer'' in both contexts, the question ``with the Federation move, will diploma issuance also be done with Federation?'' binds two different activities in one word.

From the model's side, this can be explained as ``the same function merely appearing at two different timings (issuance vs.\ session) and placements,'' and a model equipped with the three dimensions makes the underlying isomorphism explicit: Federation appearing as a ``trust layer'' in both authentication and credential presentation is the same work---as a function set of the three layers, the two planes, and the Shinken framework---done in two places with different choices along dimension 2 (timing) and dimension 1 (placement). Recognizing this, the discussion can separate ``the Federation update on the authentication side'' from ``the trust-chain design on the credential issuance/presentation side.''

The word ``Federation'' does not by itself fix a timing or a placement. OpenID Federation admits three ways of supplying the trust chain: runtime resolution by the verifier, delegation to a resolver, and supply of the chain by the entity itself at presentation. Which of these is used is a Logistics-plane choice, and the same Federation therefore occupies different positions along the three dimensions depending on it; the Logistics choices that make offline verification possible (Section~\ref{sec:II_3_2}) are available to Federation as well. Bundling moves the retrieval, not the selection of the trust anchor, which remains with the verifier.

\subsubsection{CTDL Registry: two planes bundled in one word}
The CTDL ecosystem (operated by Credential Engine) consists of three kinds of information: individual issued credentials; individual credential type definitions published on the Credential Registry~\cite{credregistry}; and the CTDL vocabulary as the language of description. A type definition is a type-level descriptive record bound to a CTID (Credential Transparency Identifier). Individual credentials pass from issuer to holder; type definitions and the vocabulary are published from the Registry and its related infrastructure.

Applied to the model: the vocabulary is ``the agreement on the data model used within the set'' (part of the policy) that the issuer set shares---content of the Constitution plane. Individual credential type definitions are likewise type-level, prior to individual acts of verification, and arrangements the community shares, so they too belong to the Constitution plane's content. The two correspond to the units of differing granularity stated in Section~\ref{sec:II_3_1}: the language for description and what is described in that language.

Here a parallelism with the trust list holds: just as the eIDAS Trusted List is Constitution-plane content that enumerates membership toward the acceptance judgment (L3), the collection of credential type definitions on the Registry is Constitution-plane content that enumerates type-level descriptions of meaning toward semantic interpretation (L2)---the form of enumeration appears in two places across layers. Indeed, separately from the Credential Registry, an Issuer Identity Registry that enumerates the issuer set has been developed through joint research by Credential Engine and the Digital Credentials Consortium, and launched by Credential Engine~\cite{iirlaunch,iirgov,iirreport,iirgf,iirlni}.\footnote{As described in the June 2025 research report~\cite{iirreport}, these registries are implemented with the protocol machinery of OpenID Federation 1.0 (entity statements and federation endpoints), yet each is configured as a single-tier structure in which the registry operator is the sole trust anchor and the subordinate listing returns a flat list of issuer DIDs. In the vocabulary of Section~\ref{sec:II_3_1}, the Constitution-plane form is therefore enumeration, with the derivation logic degenerated to list membership: the choice of protocol machinery and the choice of Constitution-plane form are independent, which is precisely the distinction the model's vocabulary makes expressible.}

By contrast, the mechanisms of the Registry, such as CTID resolution and record retrieval, are paths that deliver this content to the verifying system, and belong to the Logistics plane. A status list, by comparison, is dynamic information bound to an instance: a verification material carried by the fifth element of the Logistics plane, not content of the Constitution plane.

The interpretation from applying the model is this: the single word ``Registry'' bundles the entities of two different planes---the institution on the Constitution side and the mechanism on the Logistics side. As limited in Section~\ref{sec:II_6}, a registry in this model is the substance of a declaration and does not carry the delivery path; the CTDL Registry bundles the two under one word and one operating party, and the confusion around the word stems from that consolidation. Decompose them, and each entity settles into its own plane. Just as the eduGAIN analysis showed one mechanism (Federation) appearing at two timings and placements, here one word straddles two planes.

This decomposition also reformulates the often-noted point of single-operator operation by Credential Engine, and the reformulation splits into two observations. For CTDL and the Credential Registry, the concern stands: the role of declaring the enumeration-form Constitution content and its supply point on the Logistics side are held concurrently by a single operator, leaving the non-delegable choice of whose declaration to accept (Section~\ref{sec:II_6}) without a substantive alternative. For issuer identity, by contrast, the joint work of Credential Engine and the Digital Credentials Consortium shows that the two roles can be held by different parties: in that research the two organizations built interoperable prototype registries as distinct parties against a shared metadata and API specification, with verifier implementations consulting multiple registries and surfacing all matches~\cite{iirreport,iirgf,dcckr}. The two-plane decomposition holds that the role of declaration and the path of distribution could be carried by different parties; the CE--DCC configuration showed that the declaring role need not rest with a single operator, and single-operator concurrence is thereby shown as a design choice rather than a necessity.

Table~\ref{tab:explain} summarizes the explanatory power of the model for the cases applied in this section, contrasted with element-wise juxtaposed comparison. In this table, ``Yes'' means that the cause of the phenomenon can be identified in the vocabulary of the approach; ``Partial,'' partially; ``No,'' that the approach lacks the vocabulary to identify it.

\begin{table*}[htbp]
\caption{Explanatory power of element-wise comparison versus this model (Yes: explains; Partial: partially; No: does not explain)}
\label{tab:explain}
\centering
\footnotesize
\begin{tabular}{@{}p{2.8cm}cc p{8.6cm}@{}}
\hline\hline
\textbf{Case} & \textbf{Element-wise comparison} & \textbf{This model} & \textbf{Key point of the explanation} \\ \hline
Open Badges/CLR & Partial & Yes & The TLS chain shows the legitimacy of a domain, not legitimacy as an issuer of learner credentials (deficiency of both planes) \\
unprofiled DID/VC & Partial & Yes & A DID Method gives the resolution path (Logistics); policy, issuer set, and origin of trust (Constitution) are out of scope \\
EDC/ELM/EUDI-Wallet & Yes & Yes & Both planes are designed; offline verifiability is explained as the combination of bundling and pre-provisioning \\
CTDL Registry & No & Yes & The single word ``Registry'' bundles two planes: Constitution (governance of what is registered) and Logistics (the delivery infrastructure for resolution and retrieval) \\
eduGAIN & No & Yes & Isomorphism: the same trust-establishment function appears at two different timings (session vs.\ issuance) and placements \\
\hline
\end{tabular}
\end{table*}
The four observations of the introduction are likewise grounded in this section's analyses. That stacks do not interoperate (Observation 1) is a consequence of each stack's Constitution and Logistics planes being designed independently; the worked example of Section~\ref{sec:III_example} located where the failure arises. The tendency of discussion to center on the issuer (Observation 2) appears in the thin specification of the Logistics plane in Stacks A and B that Table~\ref{tab:stackeval} makes visible, and is paired with the polysemy of terms such as ``registry,'' which the CTDL analysis above decomposed into the two planes. The differing granularity of specification for the storage and supply of issuer information (Observation 3) can be read off Table~\ref{tab:stackeval} as a design difference in the tier at which the five elements of the Logistics plane are prescribed. And for the missing discussion of the division of the verifier's role (Observation 4), the cross-comparison along the three dimensions in Section~\ref{sec:III_5} supplied the analytic means.

\subsection{Limitations of the Evaluation}
\label{sec:III_8}
The evaluation in this section has limitations. First, the stacks and cases treated center on the learner-credential domain and reflect the authors' selection; they are not exhaustive. Second, what has been shown is interpretability for existing ecosystems; the effectiveness of using the model to design a new ecosystem is a task to be verified through future application. Third, the analysis of each stack is based on the published specifications and observable operational reality as of November 30, 2025 (due to the drafting timing of the paper~\cite{edcredi}), and may not reflect internal design intent. Fourth, the judgments in Tables~\ref{tab:stackeval} and \ref{tab:explain} were made by the authors; the agreement independent raters would reach when reclassifying with the same criteria has not been measured. Fifth, this section presents cases the model could explain; it does not conduct a systematic search for phenomena the model cannot explain (a search for counterexamples). Sixth, because judgments such as ``explicit'' and ``not specified'' take public documents as their range of evidence, they cannot distinguish what is not public from what does not exist. The literature survey and the access to the referenced specifications are as of July 11, 2026 (references added in the present revision were accessed on July 31, 2026); a referenced specification postdating the observation window is used only to characterize the current status of that specification, not to infer prior implementation or deployment intent. The specification analysis of Stack C is based on ARF v2.9.0~\cite{eudiarfv2}, the current version as of the literature-access date (see the footnote in Section~\ref{sec:III_3}).

\section{Conclusion}
\label{sec:conclusion}
This paper presented a conceptual model for organizing the siloing, the lack of interoperability, and the confusion of discussion observed in the digital credential ecosystem.

In the introduction we presented, as phenomena that the existing comparison axis of juxtaposing technical elements fails to capture, four observations together with the polysemy of terms such as registry.

In the conceptual-model section, taking the framework of the Trusted Web White Paper ver.3.0 as upstream and incorporating the refinement by the Claim Validation model and the Shinken model~\cite{claimval,shinken}, we presented a model of five functions: three layers that carry the establishment of verification (signature verification (L1), semantic interpretation (L2), validation (L3)) and two planes that carry the acquisition of verification materials (Constitution and Logistics), with the Shinken framework acting on all five to make the introduction of trust explicit. Constitution and Logistics are orthogonal, and verifiability holds only when both are present. On top of this decomposition of function we introduced a decomposition of roles---where to place each function, excluding the non-delegable selection of assumptions---along three dimensions (placement, timing, disclosure). We limited the word ``registry'' to one implementation of the Constitution plane's enumeration form, and made explicit that declaration and acceptance are separate activities. The generalization to combinations of multiple credentials requires no new layer: it is the integration of multiple sources in validation, whose point of integration becomes a new introduction of trust made explicit (\ref{sec:II_7}). From this structure we drew seven consequences, concerning the scope within which verifiability holds, origin selection, offline verifiability, and the constitution of privacy, among others (\ref{sec:II_6_conseq}).

In the application-and-evaluation section, we applied the model to the four kinds of stacks in the learner-credential domain and cross-compared existing ecosystems, including authentication systems, along the three dimensions of role decomposition. In Stacks A and B, neither plane is sufficiently prescribed; Stack C is, within the scope treated in this paper, the only example in which both planes are designed; Stack D is a structure in which a single operator holds both planes. In the cross-comparison, we organized eduGAIN, OpenID Federation, EUDI / EUDI-Wallet, and Originator Profile as differences in the choices along the three dimensions, and showed, as a demonstration, that the model makes explicit the isomorphism underlying the eduGAIN question.

We summarize the utility of the model in three points. Of these, the second has been exercised in this paper through the eduGAIN analysis; the first and third state intended uses, whose effectiveness in actual design work is a task to be verified through future application (the second limitation in Section~\ref{sec:III_8}).

\emph{A frame for evaluating proposals}: for a proposed new standard or protocol, one can ask which of the Constitution plane's constituent elements it prescribes at which tier, how it prescribes the five elements of the Logistics plane, and how it places each function along the three dimensions. A proposal that cannot answer these questions leaves a hole in the implementability of the verifying system.

\emph{The utility of terminological organization}: when terms such as ``registry,'' ``trust list,'' ``issuer metadata,'' ``Trust Framework,'' and ``Federation'' are used, making explicit which element of the Constitution or Logistics plane each denotes, and which choice along the dimensions of role decomposition it implies, enables cross-cutting discussion. The eduGAIN analysis is a concrete example of this utility.

\emph{Concretizing the verifier's viewpoint}: discussion from the verifier's viewpoint can be reduced to concrete questions---how to acquire the trust anchor to be built into the verifying system, how to gather the signed data composing the chain of trust, and where to place each function. To correct discussion biased toward the issuer's viewpoint, it is effective to keep these questions always on the table.

Premised on the model, several topics remain as future work. The first is the recursive structure of communities: a community does not exist alone but can be organized, as a part of a larger community, into a nested structure---a community within a community. The recursion appears in both planes: in the Constitution plane, the constituent elements of an upper community can be inherited by or referred to from a lower community; in the Logistics plane, a lower community can use the supply path of an upper community, or an upper community can aggregate the supply paths of lower communities. A resolution mechanism for contradictions between the policies of an upper and a lower community also belongs to this topic. The second is the design of inter-community mapping, which requires the elements of both communities' Constitution and Logistics planes to be made mutually referable, with explicit rules for how trust propagates; trust cannot be assumed to be transitive~\cite{trusttransitive}, and the machine-readable publication of mutual recognition between trust schemes~\cite{trustinterop} directly overlaps with this problem. These first two are distinct problems: the first is vertical---nesting within a line of communities---while the second is horizontal, between communities that stand as peers, and neither reduces to the other. In both, the point at which acceptance terminates remains fixed on the verifier's side. The third is principles of supply design that ease connection, such as separating the constituent elements and making role placement explicit. These, together with the correspondence of the existing ``Trust Framework'' notion to this model, are planned to be developed in separate manuscripts.

\appendices

\section{Overview of Trusted Web and the Shinken Model}
\label{app:C}
The model of this paper builds on the Trusted Web White Paper ver.3.0 (in Japanese)~\cite{tw3}, the Claim Validation model by Abe et al.~\cite{claimval}, and the development by the Shinken model (in Japanese)~\cite{shinken}. This appendix gives a brief overview of the two sources available only in Japanese; for the Claim Validation model, we refer the reader to the original paper.

The Claim Validation model and the Shinken model are described in the vocabulary of three parties: the verifier (the Validator in the former, the Verifier in the latter), the Claimant, and the Certifier. The ``issuer'' of this paper corresponds to the Certifier; the verifying party divided here into the three roles of L1--L3 corresponds to the verifier of both models.

\subsection{Trusted Web}
\label{sec:C_TW}
Trusted Web is an initiative, studied by the Trusted Web Promotion Council established under the Digital Market Competition Council of the Cabinet Secretariat, aimed at building a new framework of trust---strengthening the control of data by individuals and corporations without excessive dependence on a particular service, and making the data exchanged and the counterpart verifiable---with its root value placed on ``raising Trust by expanding the region that can be verified.'' White Paper ver.3.0~\cite{tw3} presents an architecture whose constituent elements are Verifiable Data, Verifiable Messaging, Verifiable Identity, and the Verifiable Identity community: a set of identities that, under a certain governance, shares information including the community's policy, community-specific namespaces and identifiers, a list of identities, the origin of trust, and other metadata. A community can take a recursive structure, can be established at will as long as the elements for establishment are shared, and is designed to be able to encompass what is called a Trust Framework.

\subsection{The Shinken Model}
\label{sec:C_2}
The Shinken model is an abstract model for implementing the Claim Validation model's validation as an information system (a verifying system) satisfying two requirements: determinism (the verification result for the system's input settles deterministically) and transparency (when a false positive or false negative arises, its cause can be identified). The logic of verification, however, can contain propositions not decidable from available computational evidence alone---for instance, that the certificate issuer records an authentic claim in the certificate, a matter of the issuer's operation. The Shinken model resolves this by assuming some propositions true within the logic of verification, defining this introduction of an assumption as ``trust'': the whole logic can then be constituted as decidable logic, and because the part into which trust was introduced is made explicit, transparency too is secured and one can identify which part to update when an assumption is broken.

\section*{Acknowledgment}
This paper was prepared with the assistance of generative AI. Specifically, Anthropic Claude Opus 4.6, 4.7, 4.8, 5, Fable 5, 5.1, and OpenAI GPT-5.5 Pro and GPT-5.6 Pro were used, across all sections, to assist with literature research, drafting, translation, and editorial refinement of the manuscript. The authors independently verified all technical claims and bear sole responsibility for all content of this paper.

\bibliographystyle{IEEEtran}
\bibliography{refs-en}

\begin{IEEEbiography}[{\includegraphics[width=1in,height=1.25in,clip,keepaspectratio]{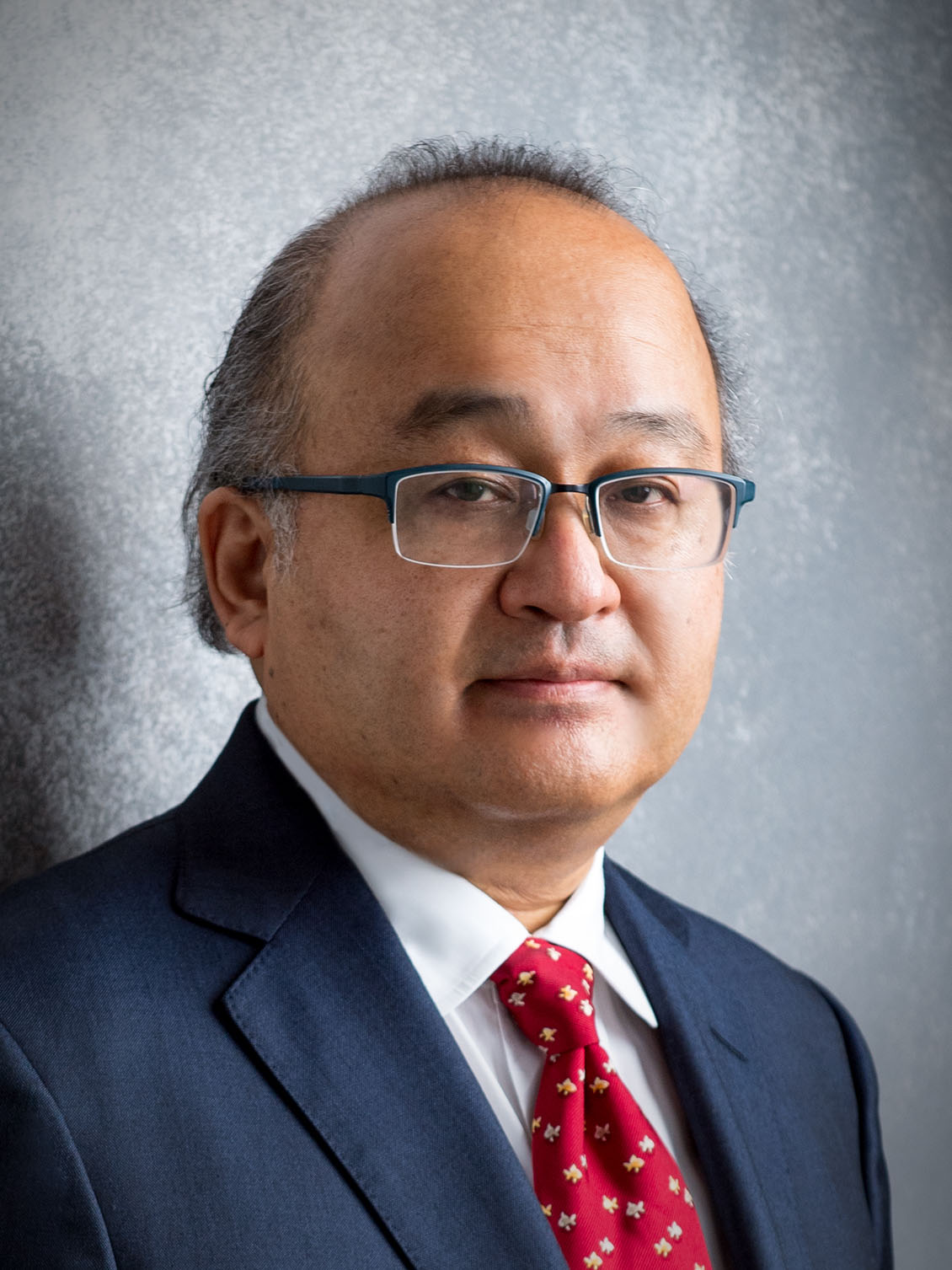}}]{Shigeya Suzuki}
(Member, IEEE) is a researcher specializing in computer networks, distributed systems, and software development. He is an active participant in the W3C Decentralized Identifier (DID) and Verifiable Credentials (VC) Working Groups, and a principal architect of the Originator Profile technology. Suzuki received his Ph.D. from the Graduate School of Media and Governance at Keio University, where he currently serves as a Project Professor at the Keio Global Research Institute. He also holds the position of Associate Director at both the Data Architecture Laboratory and the Auto-ID Laboratory within the Keio Research Institute at SFC. Suzuki is affiliated with professional organizations such as ACM, IEEE, IACR, and IPSJ, and serves on the board of the WIDE Project, which was instrumental in introducing the Internet to Japan.
\end{IEEEbiography}

\begin{IEEEbiography}[{\includegraphics[width=1in,height=1.25in,clip,keepaspectratio]{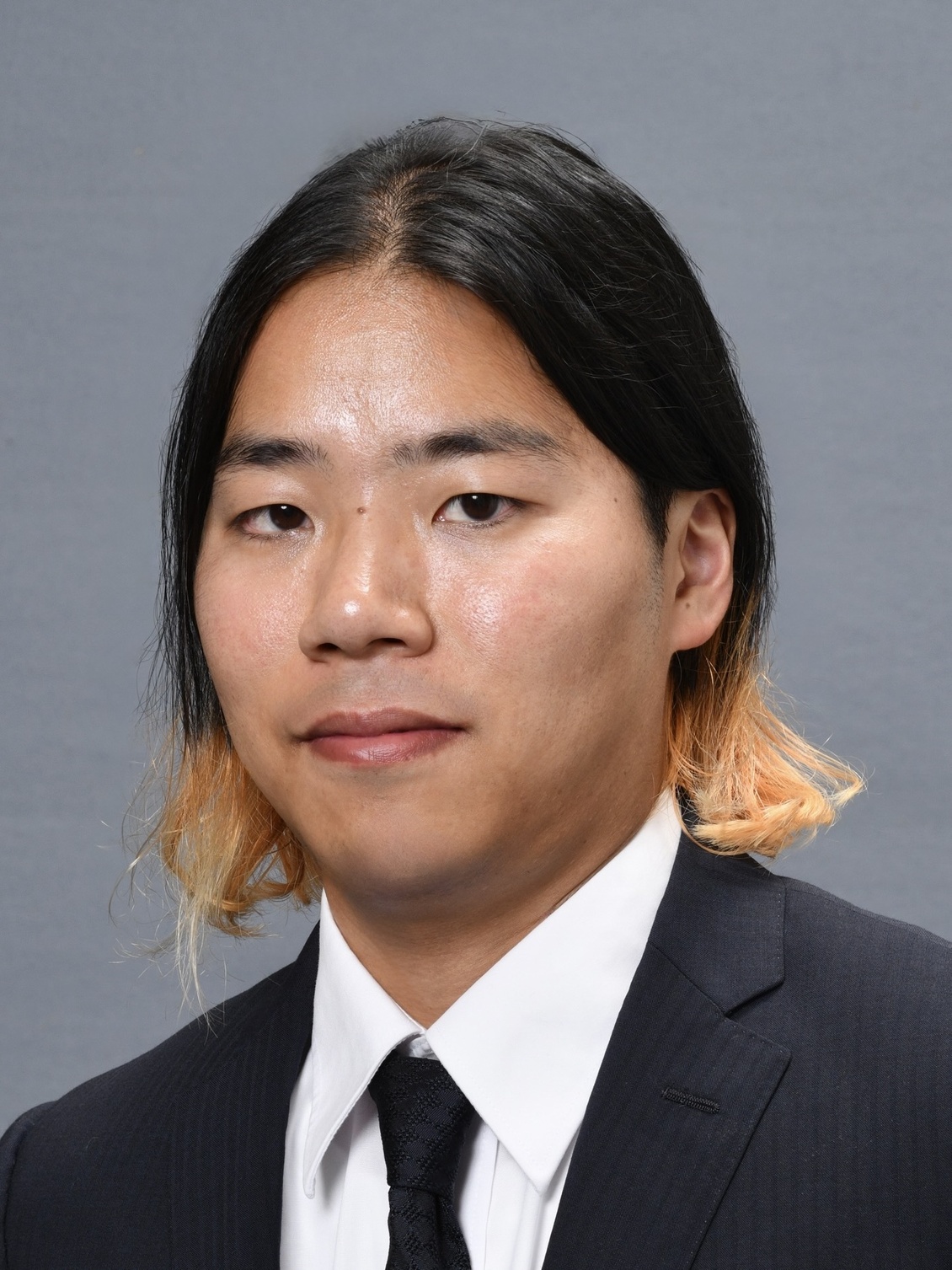}}]{Ryosuke Abe}
is an Assistant Professor at the Center for Information Infrastructure, Japan Advanced Institute of Science and Technology (JAIST), and a Project Assistant Professor at the Global Research Institute, Keio University. He received the B.A. in policy management, the M.M.G., and the Ph.D. in Media and Governance from Keio University, Kanagawa, Japan, in 2017, 2019, and 2025, respectively. From April 2022 to March 2026, he was a Project Research Associate at the Graduate School of Media and Governance, Keio University; he has held his current position since April 2026. Since 2016, he has been engaged in research on the application and infrastructure of digital certificates and blockchain. He has served on the WIDE Project Board since March 2022. He is a member of ACM and IPSJ.
\end{IEEEbiography}

\end{document}